%% file: dynamo.tex
\documentclass[fleqn,usenatbib]{mnras}

\usepackage[T1]{fontenc}

\DeclareRobustCommand{\VAN}[3]{#2}
\let\VANthebibliography\thebibliography
\def\thebibliography{\DeclareRobustCommand{\VAN}[3]{##3}\VANthebibliography}

\usepackage{graphicx}	
\usepackage{amsmath}	
\usepackage{amssymb}	

\title[Clumps in DYNAMO]{Giant Star Forming Complexes in High$-z$ Main Sequence Galaxy Analogues: The Internal Structure of Clumps in DYNAMO Galaxies}

\author[L. Lenki\'{c} et al.]{Laura Lenki\'{c},$^{1}$\thanks{Contact e-mail: \href{mailto:llenkic@umd.edu}{llenkic@umd.edu}}, Alberto D. Bolatto,$^{1,2}$ Deanne B. Fisher,$^{3,4}$ Karl Glazebrook,$^{3,4}$ Danail Obreschkow,$^{5}$ \newauthor Roberto Abraham,$^{6,7}$ and Liyualem Ambachew$^{3}$
\\
$^{1}$Department of Astronomy, University of Maryland, College Park, MD 20742, USA \\
$^{2}$Visiting Scholar at the Flatiron Institute, Center for Computational Astrophysics, New York, NY 10010, USA\\
$^{3}$Centre for Astrophysics and Supercomputing, Swinburne University of Technology, PO Box 218, Hawthorn, VIC 3122, Australia \\
$^{4}$ARC Centre of Excellence for All Sky Astrophysics in 3 Dimensions (ASTRO 3D) \\
$^{5}$International Centre for Radio Astronomy Research (ICRAR), M468, University of Western Australia, 35 Stirling Hwy., Crawley, WA 6009, Australia \\
$^{6}$David A. Dunlap Department of Astronomy and Astrophysics, University of Toronto, Toronto ON, M5S 3H4, Canada \\
$^{7}$Dunlap Institute for Astronomy and Astrophysics, University of Toronto, Toronto ON, M5S 3H4, Canada
}

\date{Accepted XXX. Received YYY; in original form ZZZ}

\pubyear{}

\begin{document}
\label{firstpage}
\pagerange{\pageref{firstpage}--\pageref{lastpage}}
\maketitle

\begin{abstract}
To indirectly study the internal structure of giant clumps in main sequence galaxies at $z \sim 1-3$, we target very turbulent and gas-rich local analogues from the DYNAMO sample with the \textit{Hubble Space Telescope}, over a wavelength range of $\sim 200-480$~nm. We present a catalog of 58 clumps identified in six DYNAMO galaxies, including the WFC3/UVIS F225W, F336W, and F467M photometry where the ($225-336$) and ($336-467$) colours are sensitive to extinction and stellar population age respectively. We measure the internal colour gradients of clumps themselves to study their age and extinction properties. We find a marked colour trend within individual clumps, where the resolved colour distributions show that clumps generally have bluer ($336-467$) colours (denoting very young ages) in their centers than at their edges, with little variation in the ($225-336$) colour associated with extinction. Furthermore, we find that clumps whose colours suggest they are older, are preferentially located closer toward the centers of their galaxies, and we find no young clumps at small galactocentric distances. Both results are consistent with simulations of high--redshift star forming systems that show clumps form via violent disc instability, and through dynamic processes migrate to the centers of their galaxies to contribute to bulge growth on timescales of a few 100 Myr, while continually forming stars in their centers. When we compare the DYNAMO clumps to those in these simulations, we find the best agreement with the long-lived clumps.
\end{abstract}




\section{Introduction} \label{sec:intro}
The morphologies of massive star-forming galaxies at $z \sim 1 - 3$ are irregular compared to local star-forming galaxies of similar mass. Rest-frame UV and optical images of these galaxies reveal giant, $\sim 0.1 - 1$ kpc sized clumps with masses ranging from $10^{7}-10^{9}$ M$_{\odot}$ \citep{forsterschreiber11,guo12,huertas-company20} that are the sites of active star formation, accounting for up to $10-20$\% of the total integrated star forming activity of galaxies at this epoch \citep{wuyts12,wuyts13}. These giant clumps are also observed in resolved molecular gas maps \citep{tacconi13} and rest-frame optical line emission spectra \citep{genzel08,genzel11}, but are less prominent in stellar mass distribution maps \citep{wuyts12}. Galaxies at this epoch are also observed to have high gas-fractions, defined as $f_{gas} = M_{H_{2}}/(M_{H_{2}}+M_{*})$ \citep[$f_{gas} \sim 0.3 - 0.8$,][]{daddi10,tacconi10,tacconi13,genzel15}, high star formation rates \citep{genzel08}, and high velocity dispersions \citep{forsterschreiber06}.

There is an ongoing debate as to the nature of these clumps and their eventual fate. Cosmological simulations \citep[e.g.,][]{dekel09,bournaud14,mandelker17} produce clumpy galaxies via cold gas accretion and gravitational disc instabilities. These clumps survive long enough in some simulations to migrate to the centers of galaxies through orbit decay by loss of orbital angular momentum, on timescales of $\lessapprox$ 500 Myr and thus contribute to galaxy bulge growth \citep{noguchi99,immeli04,bournaud07,ceverino10}. \citet{ceverino12} show in their cosmological simulations that this migration results in a clump age gradient that is much steeper than the disc age gradients \citep[see also][]{mandelker17}. Observational measurements of clump colours also provide evidence for this inward migration: redder clumps are found closer to the centers of galaxies, while bluer clumps are found preferentially at larger galactocentric distances. These galaxy-wide colour gradients are interpreted as an age gradient, where the redder clumps are thought to be older \citep{forsterschreiber11,guo12,shibuya16,soto17,guo18}. 

This scenario however, requires that clumps survive for timescales of $\sim 150$ Myr. Yet other numerical simulations (such as Feedback in Realistic Environments, FIRE; \citealt{hopkins14}, and NIHAO; \citealt{wang15}) find that clumps dissolve on timescales of $\sim 100$ Myr or less and thus do not survive long enough to contribute significantly to galaxy bulge growth \citep{buck17,oklopcic17}, even though these simulated galaxies show the same clump colour gradients as observed. Furthermore, other simulations show that increasing clump ages at small distances can be produced by inside-out formation of the disc, rather than clump migration, and that clumps disrupt on timescales of $\sim$\,50~Myr \citep{genel12}.

\citet{buck17} propose that the colour gradient found by observers may be the result of contamination from disc light. The density of disc stars is lower in the outskirts of a galaxy and higher in the center, thus the authors argue that redder stellar light in the centers of galaxies can artificially produce colour gradients such as the ones observed. Contrary to this idea, \citet{guo18} find that regardless of the chosen disc subtraction method used (ranging from no disc subtraction to very strict disc subtraction), the clump colour gradients in their sample of 1270 galaxies at $0.5 \leq z < 3.0$ are still present and only the amplitude of the gradients are affected.

Nonetheless, in agreement with observations, the NIHAO galaxies of \citet{buck17} show clumps in light maps but not in stellar mass maps. The authors use this to argue that the clumps detected in the simulated galaxies are not long-lived gravitationally bound systems that have any dynamical influence on the evolution of the disc. Rather they are simply localized (clumpy) regions of star formation. The authors furthermore postulate that dust attenuation can affect the observed structure of a galaxy by enhancing intrinsically dim clumps, and suppressing intrinsically bright ones, suggesting that clumps are simply the result of variations in extinction.

Fortunately, the existence of nearby galaxies with properties similar to those at high redshift, allows us to study the light and extinction distributions in clumpy galaxies on a resolved scale. The DYnamics of Newly Assembled Massive Objects \citep[DYNAMO;][]{green14} sample consists of very rare local ($z \sim 0.1$) galaxies whose properties are very well matched to those of main-sequence galaxies at $z\sim1-3$. \citet{green14} derive H$\alpha$ rotation curves for DYNAMO galaxies and find that the mean ionized gas velocity dispersion is $\sim 50$ km\,s$^{-1}$, reflecting much larger turbulent motions than in normal local galaxies, and infer gas fractions as high as $f_{gas} \sim 0.8$, and star formation rates in the $0.1-100$~M$_{\odot}$\,yr$^{-1}$ range. The properties of DYNAMO galaxies place them on the main sequence of star formation of $z \sim 1-2$ galaxies \citep[e.g., see Figure 1 in][]{fisher19}; as a result, we have the opportunity to investigate clump properties on resolved scales, where we can learn more about their true nature and evolutionary fate.

In this work, we investigate the internal radial variation of age and extinction of individual clumps in six DYNAMO galaxies through colour measurements, as well as galaxy-wide radial trends in colour and age evolution. The paper is structured as follows: \S\ref{sec:obs} describes the targets and their HST observations, \S\ref{sec:analysis} describes our clump identification process, point-spread-function matching procedure and other data analysis, in \S\ref{sec:results} we present and discuss our photometry results, finally we summarize and conclude our work in \S\ref{sec:con}. 

We assume $\Lambda$CDM cosmology with $H_{0} = 70$ km\,s$^{-1}$\,Mpc$^{-1}$, $\Omega_{m} = 0.3$, and $\Omega_{\Lambda} = 0.7$, and a Kroupa initial mass function \citep[IMF;][]{kroupa01} throughout this paper. All magnitudes we report are AB magnitudes. The physical scale corresponding to the typical redshift of our sources ($z \sim 0.1$) is 1.844 kpc per arcsecond \citep{wright06}.

\section{Observations} \label{sec:obs}
\subsection{Sample} \label{subsec: gal_props}
The DYnamics of Newly Assembled Massive Objects \citep[DYNAMO;][]{green14} sample consists of 67 galaxies selected from the Sloan Digital Sky Survey DR4 \citep[SDSS;][]{adelman-mccarthy16} to cover a range of H$\alpha$ luminosities up to L$_{\mathrm{H_{\alpha}}} > 10^{42}$ erg\,s$^{-1}$, in the SDSS 3.0\arcsec\, diameter fiber, and to lie in two redshift windows to avoid atmospheric absorption of H$_{\alpha}$ (centered at $z \sim 0.075$ and $z \sim 0.13$). DYNAMO galaxies span stellar masses from $10^{9} - 10^{11}$ M$_{\odot}$ and SFRs from $\sim 0.1 - 100$ M$_{\odot}$\,yr$^{-1}$.

In this work, we consider a sub-sample of six DYNAMO galaxies selected to build on a multi-wavelength study of these objects: DYNAMO D13-5, D15-3, G04-1, G08-5, G14-1, and G20-2, which are classified as rotating discs based on H$\alpha$ kinematics. Galaxies G04-1, G14-1, and G20-2 are furthermore classified as ``compact'' rotating discs \citep{green14}. These galaxies are ones for whom the SDSS $r-$band exponential scale lengths are less than 3~kpc. Since the resolution for these objects is poorer, their kinematic classifications are less reliable. All six of these objects have molecular gas measurements from CO($1-0$) observations with the Plateau de Bure Interferometer (PdBI), and have gas fractions of $20-30$\% and depletion timescales of $t_{dep} \sim 0.5$ Gyr \citep{fisher14,white17}. Using HST H$\alpha$ observations and Keck/OSIRIS Pa$\alpha$ observations, \citet{bassett17} created spatially resolved ($\sim 0.8-1$ kpc) H$\alpha$ extinction maps for galaxies D13-5, G04-1, and G20-2, showing that there is only mild spatial variability in the amount of extinction within a given galaxy, and finding no evidence for highly attenuated star-forming clumps. Finally, the MPA-JHU Value Added Catalog\footnote{\href{https://wwwmpa.mpa-garching.mpg.de/SDSS/DR4/}{https://wwwmpa.mpa-garching.mpg.de/SDSS/DR4/}} provides gas-phase metallicity measurements for $\sim$\,53,000 star forming galaxies, including the six DYNAMO galaxies we study here, derived from comparing several strong emission lines ([OII]$\lambda$3727, H$\beta \lambda$4861, [OIII]$\lambda$5007, H$\alpha$, [NII]$\lambda$6584, [SII]$\lambda$6717, and [SII]$\lambda$6731) to photoionization models \citep[see][]{tremonti04}. These properties are summarized in Table \ref{tab:properties}.

\input{properties}

\subsection{HST Observations} \label{subsec: obs}
\begin{figure*}
    \centering
    \includegraphics[width=\textwidth]{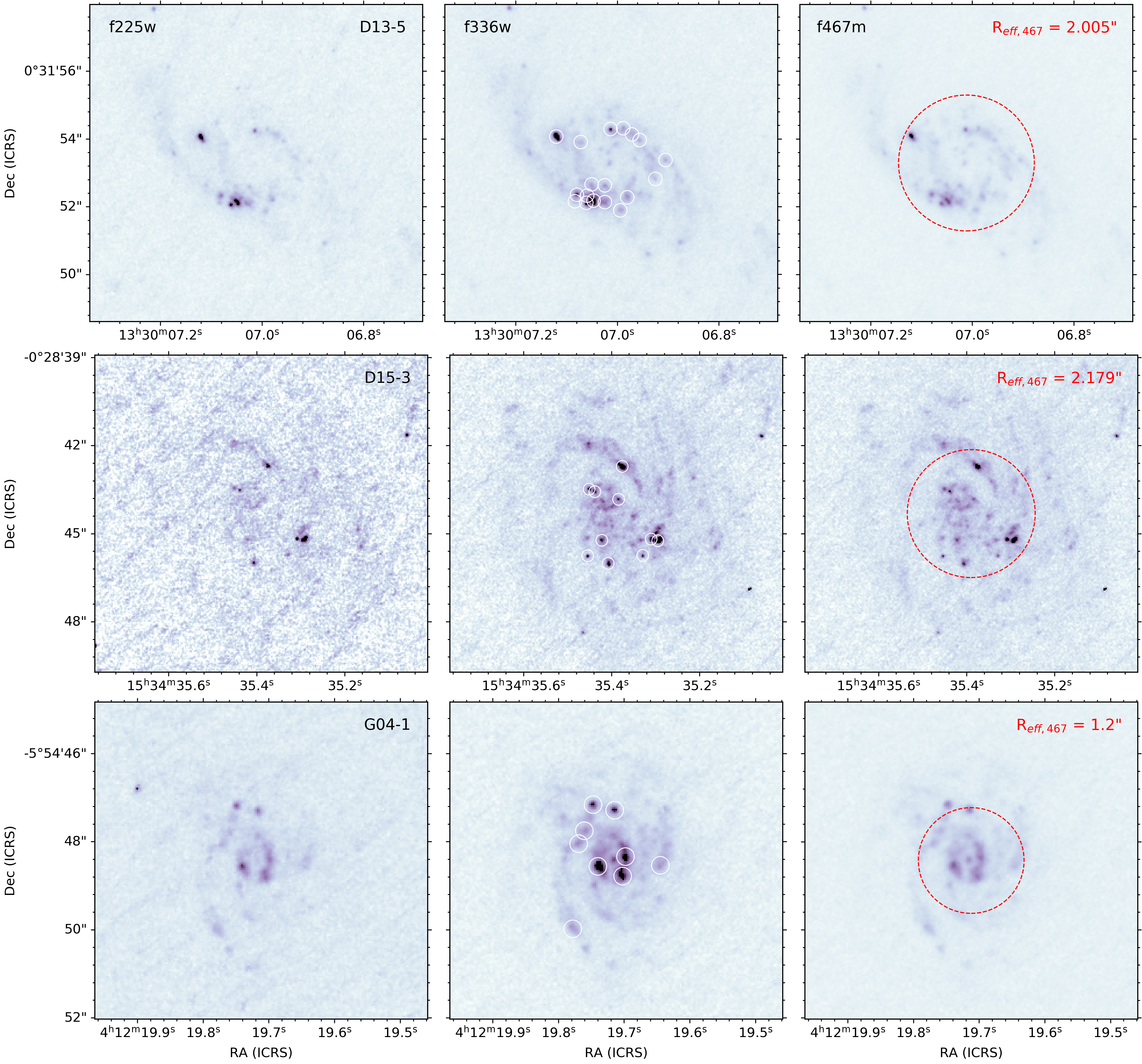}
    \caption{HST observations of galaxies D13-5, D15-3, and G04-1: F225W (left), F336W (middle), F467M (right). The F467M measured effective radius of each galaxy is shown in the right panels (dashed red circle). The middle panels outline in white each detected clump, as described in \S\ref{sec:analysis}. Galaxy G04-1 is classified as a ``compact'' rotating disc \citep{green14}.}
    \label{fig:obs1}
\end{figure*}

\begin{figure*}
    \centering
    \includegraphics[width=\textwidth]{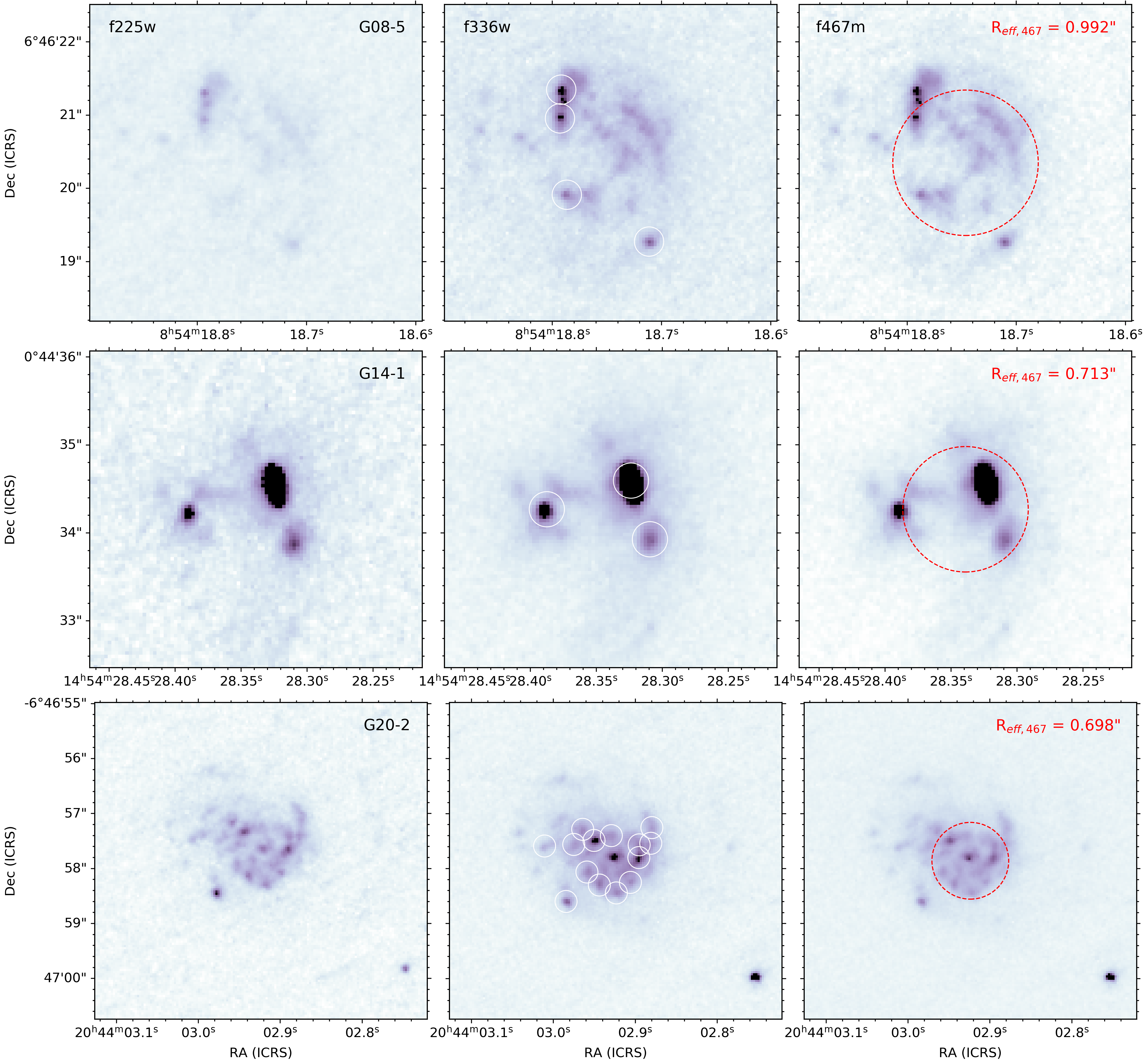}
    \caption{Same as Figure \ref{fig:obs1}, but now showing galaxies G08-5, G14-1, and G20-2. Galaxies G14-1 and G20-2 are classified as ``compact'' rotating discs; therefore, their classification may be less reliable \citep{green14}.}
    \label{fig:obs2}
\end{figure*}

The six DYNAMO galaxies in our sub-sample were observed with the \textit{Hubble Space Telescope} (HST) Wide Field Camera 3 (WFC3) UVIS channel in three filters: F225W (UV wide), F336W (U), and F467M (Str\"{o}mgren b). Observations were taken between February and November 2018. The full-width-half-maximum (FWHM) of the point-spread-function (PSF) of each filter is: $\sim$0.092\arcsec\, for F225W, $\sim$0.080\arcsec\, for F336W and F467M. We note that the observations are not diffraction limited until longward of 500 nm, thus the broader F225W PSF at shorter wavelengths. We present these data in Figures \ref{fig:obs1} and \ref{fig:obs2}.

\section{Data Analysis} \label{sec:analysis}
\subsection{Kernel Generation for PSF Matching} \label{subsec: psf_match}
The goal of this work is to measure the internal colours of clumps in DYNAMO galaxies, taking full advantage of the HST resolution. To do this, it is necessary to transform images from different filters to a common point spread function (PSF). Since the HST filters we use here do not have the same PSFs, the variation of structure on different spatial scales will result in unphysical colours when directly comparing images with different PSFs \citep[see e.g.,][]{gordon08,aniano11}. To transform images to a common PSF, we need to generate a convolution kernel, such that when the PSF of one filter (F336W or F467M) is convolved with this kernel, it will resemble the broader PSF of the second filter (F225W):

\begin{equation}
    \Psi_{225} = \Psi_{336} \circledast K\{336 \rightarrow 225\} \label{eqn:conv}
\end{equation}

\noindent where $\Psi$ represents the filter PSFs, and $K$ represents the convolution kernel. The convolution kernel that will satisfy this equation can be generated by taking the Fourier Transform (FT) of the same equation:

\begin{align}
    \mathrm{FT}(\Psi_{225}) &= \mathrm{FT}(\Psi_{336} \circledast K\{336 \rightarrow 225\}) \newline \\
    &= \mathrm{FT}(\Psi_{336}) \times \mathrm{FT}(K\{336 \rightarrow 225\}),
\end{align}

\noindent solving for the kernel, and taking an inverse Fourier Transform, we find that:

\begin{equation} \label{eq:kernel}
    K\{336 \rightarrow 225\} = \mathrm{FT}^{-1}\left[\mathrm{FT}(\Psi_{225})\times\frac{1}{\mathrm{FT}(\Psi_{336})}\right].
\end{equation}

\noindent In reality, computing the convolution kernel numerically requires additional steps to ensure that it is stable and performs well. We follow the steps to do this as outlined in detail in \citet{aniano11} and summarize them here for completeness.

\subsubsection{HST PSFs} \label{subsubsec: psfs}
We obtained the PSFs for the F225W, F336W, F467M filters from the \texttt{Tiny Tim}\footnote{https://www.stsci.edu/hst/instrumentation/focus-and-pointing/focus/tiny-tim-hst-psf-modeling} tool, version 7.5, by downloading and installing the source code available on \texttt{GitHub}. The three PSFs are 89$\times$89 pixels and have a pixel scale of 0.04\arcsec\,, matching the pixel scale of the HST images; therefore, we perform no gridding, padding or centering of the PSFs.

\subsubsection{Circularizing the PSFs} \label{subsubsec: circularize}
The PSFs modeled by \texttt{Tiny Tim} are not rotationally symmetric; therefore, the first step we take is to circularize them. Following \citet{aniano11}, we circularize the PSFs by performing 14 rotations of the PSF, and averaging after each rotation, such that the final PSF is invariant for any rotation that is a multiple of 360$^{\circ}$/$2^{14} = 0.022^{\circ}$. We begin by rotating the original PSF, $\Psi$, by the smallest angle first: $\theta_{1} = 0.022^{\circ}$ to produce a new PSF image, $\Psi'$, and then compute the average of the two: $\overline{\Psi} = \frac{1}{2} \times (\Psi + \Psi')$. We then rotate $\overline{\Psi}$ by $\theta_{2} = $ 360$^{\circ}$/$2^{13}$ and repeat the process until $\theta_{14} = $ 360$^{\circ}$/$2^{1}$. 

\subsubsection{Compute the Kernels} \label{subsubsec: kernel}
We compute the Fourier Transform of each circularized PSF using the \texttt{Python numpy} two-dimensional Fast Fourier Transform (FFT) function \texttt{fft.fft2}. We compute the $K\{336 \rightarrow 225\}$ and $K\{467 \rightarrow 225\}$ kernels according to equation \ref{eq:kernel}, by evaluating the $1/FT(\Psi_{336})$ and $1/FT(\Psi_{467})$ terms where they are not equal to zero. After taking the inverse Fourier Transform of the kernel, we normalize the integral of the kernels to unity to ensure flux conservation (see also \texttt{photutils.psf.matching}\footnote{https://photutils.readthedocs.io/en/stable/psf\_matching.html}).

\subsubsection{Kernel Testing} \label{subsubsec: kernel_test}
\begin{figure*}
    \centering
    \includegraphics[width=\textwidth]{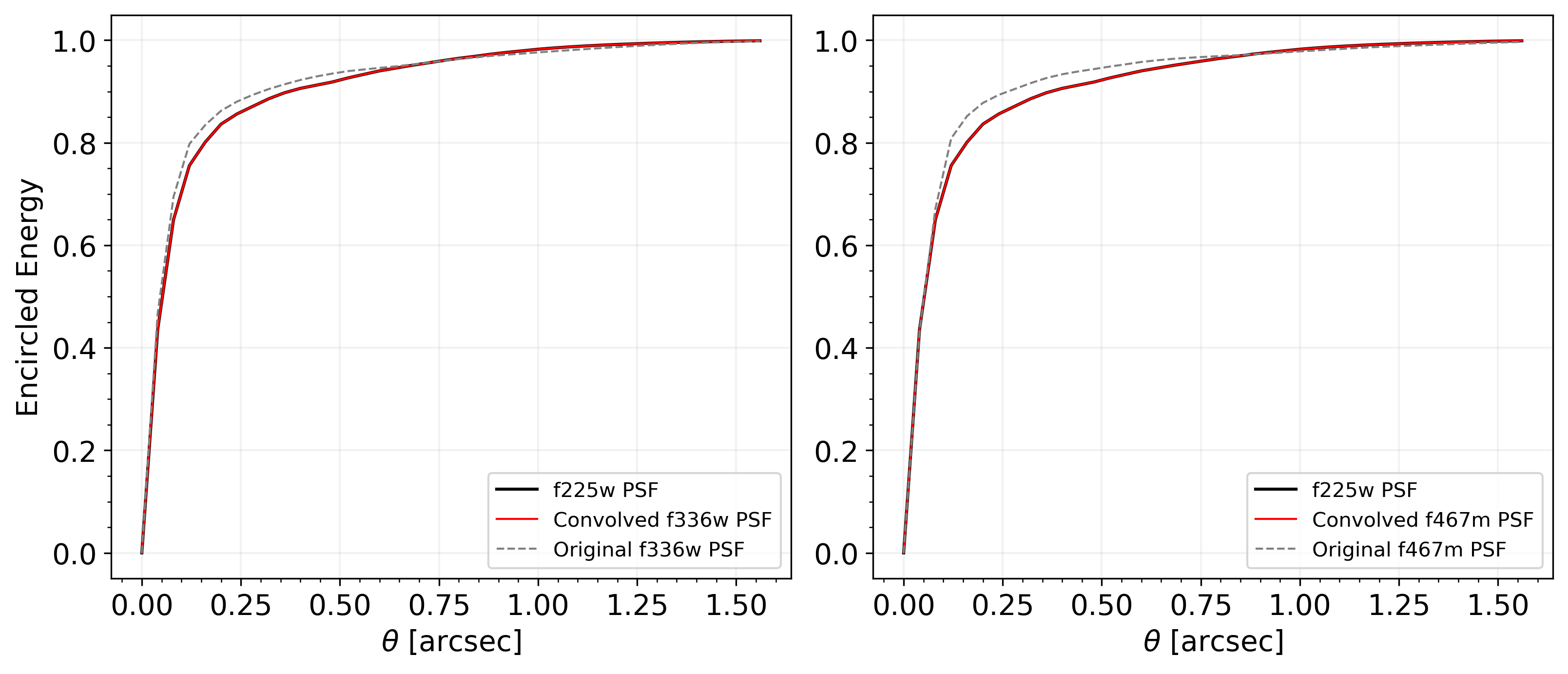}
    \caption{A comparison of the encircled energy (energy with an aperture of radius $\theta$ divided by the total energy) of the target PSF, F225W, and the source PSFs: F336W and F467M. The solid black lines are the encircled energy of the F225W PSF, while the dashed gray lines and and solid red lines are the original and convolved F336W (left) and F467M (right) PSFs. Our convolution kernels perform well and are capable of matching the target PSF.}
    \label{fig:encircled_energy}
\end{figure*}

We test our kernels according to the two metrics defined by \citet{aniano11}: measuring how accurately the kernel redistributes flux ($D$, their equation 20), and studying the negative values in each kernel ($W_{-}$, their equation 21), and by comparing the encircled energy of the target PSF (F225W) and the convolved source PSFs (F336W and F467M). We show the latter in Figure \ref{fig:encircled_energy}, where the black line is the encircled energy of the F225W PSF, the dashed line is the original source PSF, and red is the convolved source PSF (left: F336W, right: F467M). We see from this that our PSF-matching reproduces well the target F225W PSF. The integral of the absolute value of the difference between the target PSF and the PSF derived from the convolution kernel is the definition of $D$, and $D = 0$ for perfect kernel performance. For the $K\{336 \rightarrow 225\}$ and $K\{467 \rightarrow 225\}$ kernels we derive, $D_{336} = 0.0.00015$ and $D_{467} = 0.00016$ respectively, which indicate very good agreement between the PSFs. The $W_{-}$ value is defined as the integral of the negative values in the kernel, and a kernel with a large value of $W_{-}$ may amplify image artifacts. \citeauthor{aniano11} find from experimenting with several kernels that a value of $W_{-} \sim 0.3$ corresponds to a kernel that is very safe to use. Our kernels have $W_{-,336} = 0.01$ and $W_{-,467} = 0.21$, thus we conclude that they are safe to use. 

\subsection{Clump Selection} \label{subsec: clumpd_ID}
\begin{figure*}
    \centering
    \includegraphics[width=\textwidth]{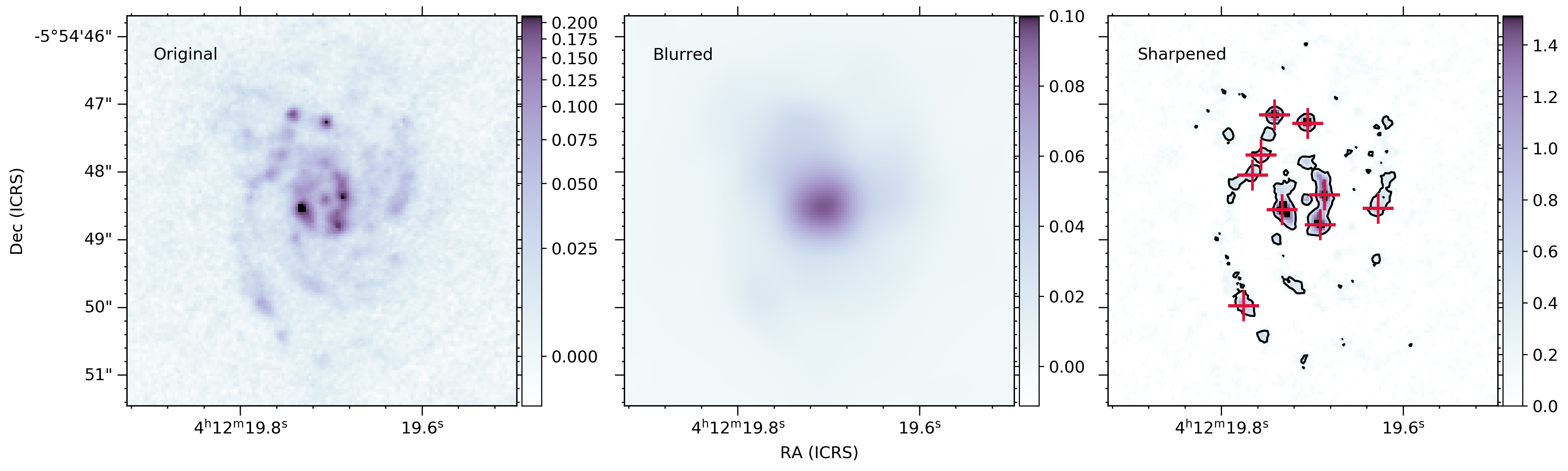}
    \caption{Outline of the unsharp masking technique for identifying clumps in DYNAMO galaxies: original F336W image of the galaxy G04-1 (left), the F336W image smoothed by convolving with a Gaussian kernel with a 15 pixel (0.6\arcsec) FWHM (middle), the contrast image that is created by "blending'' the original and smoothed image (right). The black contours outline regions that are at 5$\times$ the standard deviation of the image, while the crimson coloured crosses mark the central positions of the clumps that are identified in this galaxy and are selected to be brighter than 5$\times$ the standard deviation of the image and to have a minimum diameter of 5 pixels (0.2\arcsec).}
    \label{fig:unsharp_mask}
\end{figure*}

To systematically select clumps in our sample of galaxies, we use the unsharp masking technique, after which we run a source detection algorithm to identify clumps. The unsharp masking technique is a filter that removes low-frequency components of an image, thereby sharpening said image. Here we describe our procedure, and refer to Figure \ref{fig:unsharp_mask} where we use galaxy G04-1 as an example to illustrate this. We identify clumps from the F336W images (left panel of Figure \ref{fig:unsharp_mask}) by first convolving each image with a Gaussian kernel whose full-width-half-maximum (FWHM) is 15 pixels, which at the pixel scale of the image (0.04\arcsec/pixel) corresponds to 0.6\arcsec, or 7.5$\times$ the size of the PSF for this filter. The result of this convolution is a smoothed version of the F336W image, shown in the middle panel of Figure \ref{fig:unsharp_mask}. Finally a sharpened (contrast) image is produced by blending the original image with the smoothed image (right panel of Figure \ref{fig:unsharp_mask}):

\begin{equation}
    \mathrm{sharpened}\, =\, \mathrm{original}\, +\, (\mathrm{original} - \mathrm{blurred})\,\times\, C
\end{equation}

\noindent where $C$ is a constant that controls how much lighter and darker edge borders become; we use a value of 15 for all galaxies in our sample. We then use these sharpened images to identify clumps in each of the six galaxies.

To detect clumps we use the blob detection algorithm based on the Laplacian of a Gaussian implemented in the \texttt{Python} package \texttt{scikit-image} \citep[see \texttt{skimage.feature.blob\_log};][]{vanDerWalt14}. This algorithm works by convolving the image with a Gaussian kernel of a given size, then applying the Laplacian operator to the convolved image, which results in a strong response for clumps ("blobs``) on a given background. The response of the operator is however sensitive to the relation between the size of the smoothing kernel, and the size of the clump. For this reason, the \texttt{scikit-image} implementation of this algorithm, the \texttt{blob\_log} function, takes as input the minimum and maximum standard deviation to use for the convolution kernels, and the number of kernels to test between the minimum and maximum. 

We impose the criterion that clumps must have a minimum diameter of 2.5$\times$ the size of the F336W PSF. The F336W PSF has a size of 2 pixels, thus all our clumps are required to have a minimum diameter of 5 pixels. We therefore input into the blob detector function a minimum standard deviation of $2.5/\sqrt{2}$, and test smoothing kernels that increase in units of 0.05, up to a maximum of a 9 pixel diameter. The clump finder also takes as input a threshold parameter which controls the required intensity a clump needs to have to be detected. We set our detection limit such that clumps are regions that are 5$\times$ the standard deviation of the sharpened image (see black contours in the right panel of Figure \ref{fig:unsharp_mask}). The clumps that meet all these requirements are indicated by crimson coloured crosses in the right panel of Figure \ref{fig:unsharp_mask}.

\citet{fisher17a} finds that DYNAMO clumps have a median diameter of $\sim$\,450~pc in HST H$\alpha$ images, in two ways: (1) by fitting 2D Gaussian functions to each clump, then defining the clump size as the geometric mean of the major ($a$) and minor ($b$) axes resulting from the 2D Gaussian fit: 2\,$\times\,\sqrt{a \times b}$, and (2) by fitting a 1D Gaussian to the light profile of each clump, then defining the clump size as 2\,$\times\sigma$ from the 1D Gaussian fits. We adopt the latter technique: for each clump, we mask neighboring clumps, extract the clump light profile, fit a 1D Gaussian, then define the clump size as $r_{clump} = 2\times\sigma$, which encompasses 95\% of the clump light. We report these sizes in Table \ref{tab:measurements} in Appendix \ref{app:measurements}. From this, we find a median clump diameter of $\sim$\,400~pc, similar to the \citet{fisher17a} results. 

\subsection{Photometry} \label{subsec: photometry}
The motivation for the filter choice to observe the six DYNAMO galaxies in our sample was the fact that they provide colours that are sensitive to stellar population age and extinction. At the DYNAMO redshifts, the F225W and F336W filters bracket the UV slope, thus the $225 - 336$ colour is sensitive to changes in dust extinction. Conversely, the F336W and F467M filters measure the 4000\AA\, break, thus they bracket stellar emission and the $336 - 467$ colour is sensitive to changes in the age of the underlying stellar population. With these two colour combinations, we are able to establish constraints on how much extinction and age each contribute to the evolution of the clump colours. To this end, we modified \texttt{Starburst99} \citep{leitherer99} to use the \textit{HST} filter transmission curves for determining colours, and to blueshift by $1 + z$ the transmission curves of the standard HST filters to derive the colours at the rest wavelengths of interest. We then derive models for a single burst of star formation at solar metallicity to track how the two colours we measure here evolve with time, in steps of 0.1~Myr. As we perform our photometry and derive results from them, we will compare our colours to those of \texttt{Starburst99} to infer the age and extinction structures within clumps.

\subsubsection{Aperture Photometry} \label{subsubsec: ap_phot}
We perform aperture photometry for each clump identified by the procedure described in the previous section. We center a circular aperture on every clump with radius equal to $r_{clump}$, and define a circular annulus whose inner radius is 4$\times$$r_{clump}$, and whose outer radius is 5$\times$$r_{clump}$, for disc subtraction. We mask from this annulus any pixels that are contaminated by other nearby clumps. We then calculate the clumps flux as:

\begin{equation}
    F = \sum_{i = 1}^{n_{A}}F_{i} - n_{A}\overline{B}
\end{equation}

\noindent where $F$ is the total flux in electrons per second, $n_{A}$ is the total number of pixels in the clump aperture, and $\overline{B}$ is the median disc light measured in the disc annulus. We perform the above calculation with and without disc subtraction for comparison.

We estimate the uncertainties on all of our measurements by calculating the aperture signal$-$to$-$noise ratio (S/N) using the CCD equation \citep[see for e.g.,][]{howell06}:

\begin{equation} \label{eq:snr}
    \frac{\mathrm{S}}{\mathrm{N}} = \frac{N_{*}t}{\sqrt{N_{*}t + n_{A}(1 + \frac{n_{A}}{n_{B}})(N_{S}t + N_{D}t + N_{R}^{2} + G^{2}\sigma_{f}^{2})}}
\end{equation}

\noindent where $N_{*}t$ is the total aperture counts (i.e., counts s$^{-1} \times$ exposure time), $n_{A}$ is the total number of pixels in the clump aperture, $n_{B}$ is the total number of pixels in the background annulus, $N_{S}t$ is the total counts per pixel from the background, $N_{D}t$ is the total number of counts due to the dark current, $N_{R}^{2}$ is the total number of counts per pixel from the read noise, and $G^{2}\sigma_{f}^{2}$ accounts for the error introduced by the digitization noise within the A/D converter. These terms are obtained in the following way: (1) $N_{*}t$ and $N_{S}t$ are what we measure, (2) we use a value of $\sim$8e$^{-}$/hr/pixel for the WFC3 CCD dark current to calculate $N_{D}t$ (late-2017 value from the WFC3 Data Handbook, version 4.0 -- May 2018\footnote{https://hst-docs.stsci.edu/wfc3dhb}), and (3) we obtain the read noise and gain terms from the images headers. Since we report results in magnitudes, the final uncertainties are:

\begin{equation} \label{eq:mags}
    \sigma_{magnitudes} = \frac{1.0857}{\mathrm{S/N}}.
\end{equation}

We perform the above calculations with and without disc-subtraction, and record the non-disc-subtracted integrated measurements along with their uncertainties in Appendix \ref{app:measurements}, Table \ref{tab:measurements}. We choose to exclude disc subtraction, because as we will show in \S\ref{subsubsec: gradients}, the disc colour distributions are relatively flat, thus including disc subtraction will only increase the uncertainties of our measurements.

\begin{figure}
    \centering
    \includegraphics[width=\columnwidth]{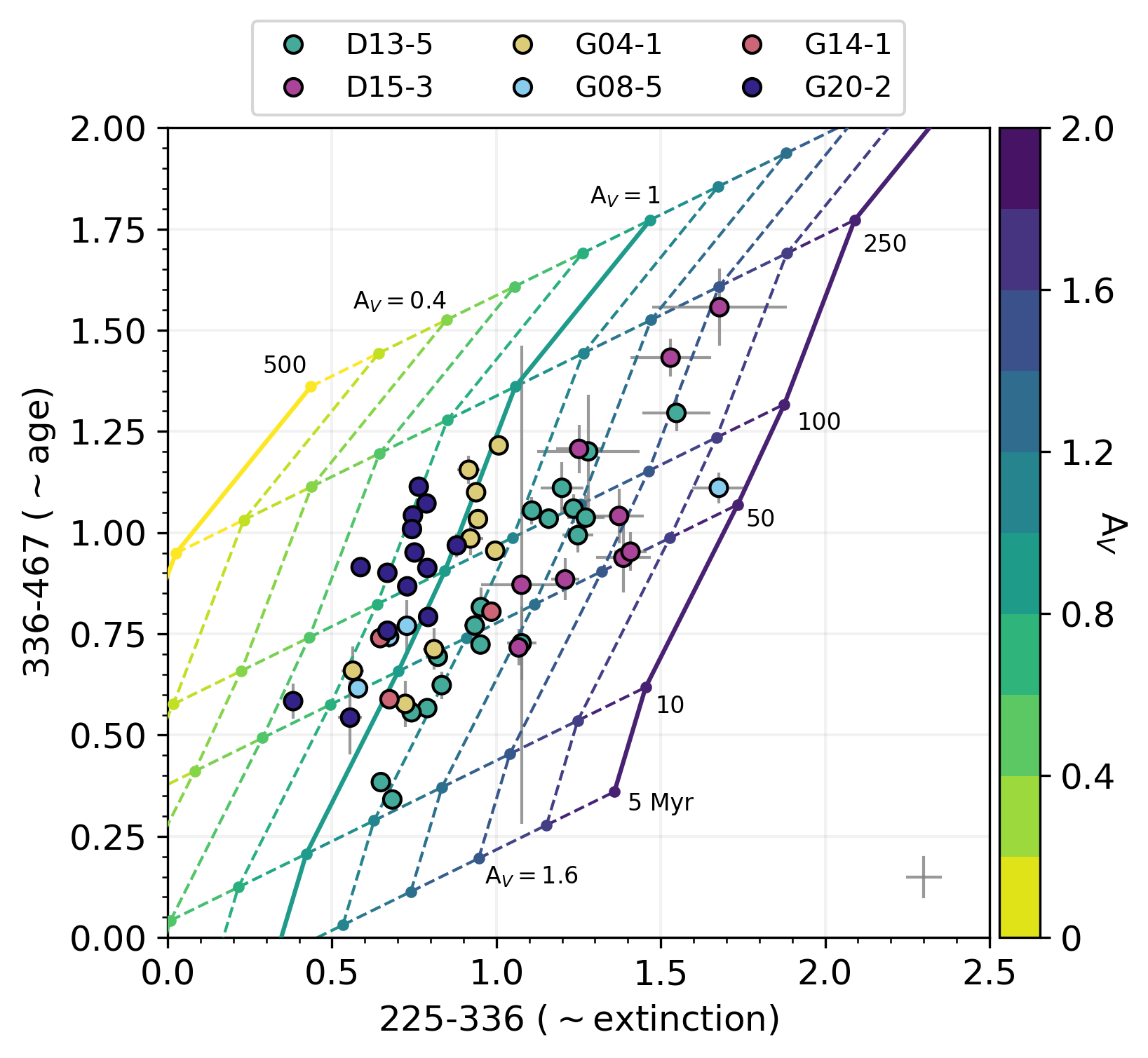}
    \caption{colour-colour diagram of the $336-467$ colour (proxy for age) versus the $225-336$ colour (proxy for extinction), showing the integrated colours of each clump in our sample. The grey error bars are the photometric uncertainties, and we show the median errors bars in the bottom right corner. The coloured lines represent the evolution of these two colours as modeled by \texttt{Starburst99}, with extinction applied ranging from $A_{V} = 0$ to $A_{V} = 2$: the solid coloured lines are $A_{V} = 0, 1, 2$, while the dashed lines increase $A_{V}$ in steps of 0.2 ($A_{V} = 0.4, 1, 1.6$ are labeled). The points on each line correspond to specific times after a single burst of star formation: 5, 10, 50, 100, 250, 500 Myr (labeled). These colours suggest clump ages as young as 10~Myr, and no older than 250~Myr, consistent with the ages derived in high$-z$ simulations of clumpy star forming galaxies \citep{bournaud14}. They also suggest visual band extinction values in the range $A_{V} \sim 0.6 - 2.0$, while for any one given galaxy, extinction does not vary drastically from clump-to-clump. This is consistent with the findings of \citet{bassett17}; in addition, we derive $A_{V}$ clump values that are in agreement with theirs for our overlapping galaxies (D13-5, G04-1, G20-2).}
    \label{fig:colour_colour}
\end{figure}

We plot the non-disc-subtracted integrated colours in Figure \ref{fig:colour_colour}, where each set of coloured points corresponds to one of the six galaxies in our sample. The solid and dashed coloured grid overlaid in the background represent the change of the $225-336$ and $336-467$ colours, as derived by \texttt{Starburst99} for a single burst of star formation and solar metallicity, and for varying degrees of extinction. We use the \citet{cardelli89} extinction law to derive a relation between the $225-336$ and $336-467$ colours and $A_{v}$, the extinction at rest-frame V-band. We then begin with the \texttt{Starburst99} model with an $A_{v} = 0$ magnitudes and then redden the model by increasing $A_{v}$ in steps of $0.2$ magnitudes, until we reach $A_{v} = 2$ magnitudes. Thus, each individual coloured line represents the evolution of the $225-336$ and $336-467$ colours at a fixed value of $A_{v}$, for ages ranging from 5 Myr to 500 Myr. 

We can see from this that dust extinction primarily moves the colours horizontally while age primarily moves the colours vertically. Though there is some diagonal evolution with increasing extinction and age, indicating that both affect the colours, it is clear that (by experimental design) a change in extinction has a much larger effect on the $225-336$ colour, while a change in age has a much larger effect on the $336-467$ colour. The integrated colours of the clumps in our DYNAMO galaxies all lie within ages of $10 - 250$ Myr with $A_{v}$ extinction ranging from 0.6 to 2.0 magnitudes. To assess the impact of our choice of metallicity on the ages we infer, we compare the model we adopt here (single burst of star formation and solar metallicity), to a model with the same star formation history and 40\% solar metallicity in Figure \ref{fig:model-comp} in Appendix \ref{app:model-comp}. This changes the ages we infer from our current adopted model by a factor of $\sim$\,1.4; namely clumps appear older by a factor of $\sim$\,1.4 in the lower metallicity model. Assuming a continuous star formation history rather than a burst of star formation has a much larger impact on the inferred ages; in this case the inferred ages increase by several factors of 10, with clump ages $\sim$\,2~Gyr in some cases, for both the solar metallicity and 40\% solar metallicity scenarios. We note, however, that these models likely represent a simplified scenario, and that the true star formation history of these clumps may be a combination of both bursty and continuous star formation.

\subsubsection{Surface Brightness Profiles} \label{subsubsec:sbps}
\begin{figure*}
    \centering
    \includegraphics[width=\textwidth]{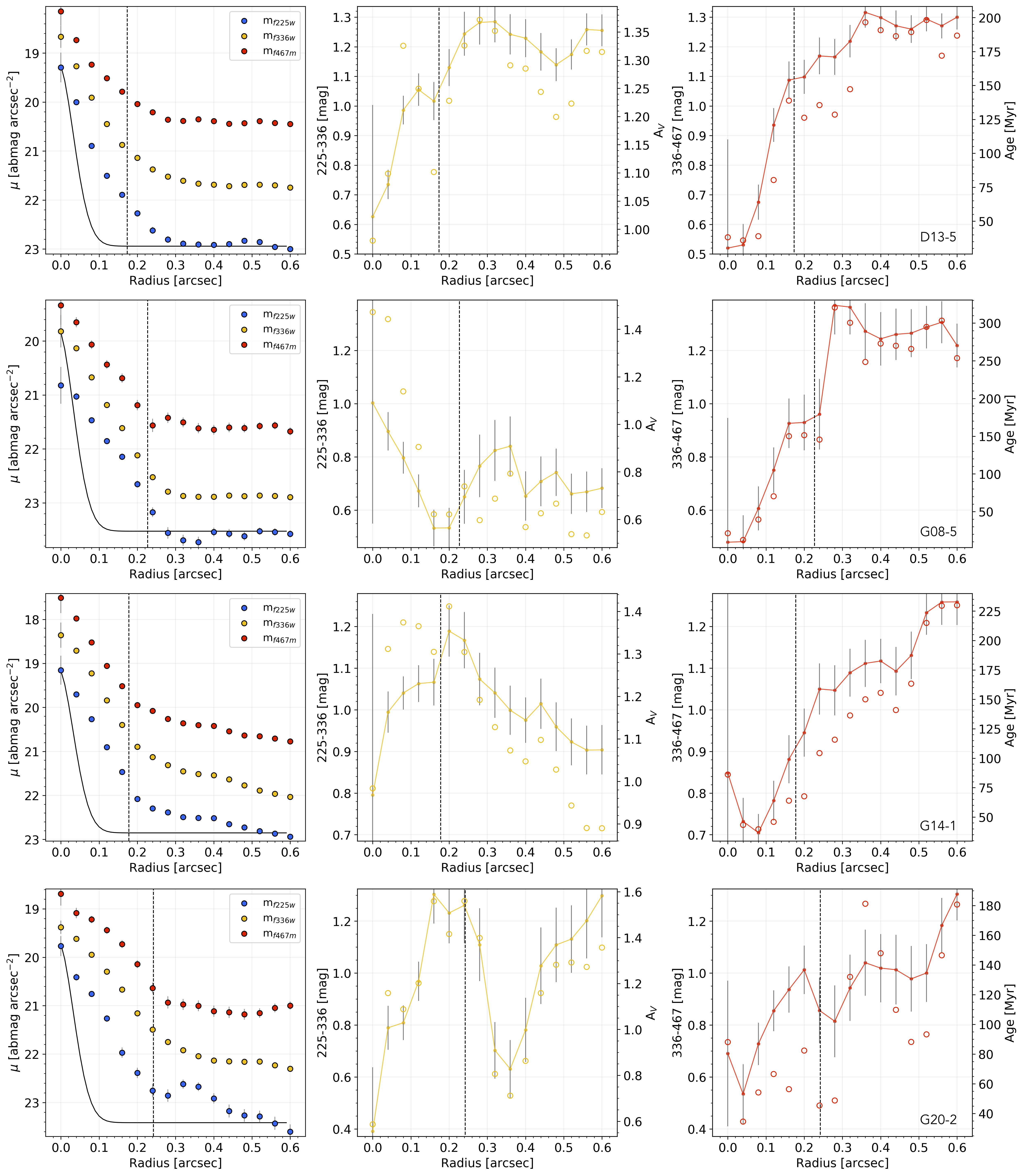}
    \caption{\textit{Left:} The $\mu_{f225w}$ (blue circles), $\mu_{f336w}$ (yellow circles), and $\mu_{f467m}$ (red circles) surface brightness profiles as a function of distance from the clump center, for clumps in galaxies D13-5 (clump 4), G08-5 (clump 3), G14-1 (clump 3), and G20-2 (clump 9). \textit{Middle:} The $225-336$ colour profile (yellow filled circles) and the conversion of these colours to $A_{V}$ extinction (open yellow circles). \textit{Right:} The $336-467$ colour profile (red filled circles) and the conversion to age (open red circles). We plot the uncertainties as gray lines, the dashed vertical lines indicate $r_{clump}$, and the solid black curve is the F225W PSF. When we map these colours to age and extinction, we see that there are clumps that show clear increase in age from the center and outward, such as we show in galaxy D13-5 and G08-5, while other clumps show a large increase in extinction rather than age, such as we show in galaxy G14-1 and G20-2.}
    \label{fig:g04-1-c2}
\end{figure*}

In addition to measuring integrated fluxes and colours for each clump in our sample, we also derive surface brightness profiles as a means of investigating clump properties as a function of distance from the clump center. We use a set of concentric circular annuli, centered on each clump, where each annulus is two pixels wide (0.08\arcsec, or approximately 150~pc for $z \sim 0.1$), and overlaps with the previous annulus by one pixel (therefore the central radius of the annulus moves by $0.04\arcsec$ for each measurement). We do this to smooth the surface brightness profiles because each clump is only a few pixels across, and we are therefore working with small numbers of pixels. We sum the signal within the annulus, determine its area, and then calculate the surface brightness according to:

\begin{equation}
    \mu \:[\mathrm{mag\,arcsec^{-2}}] = m_{HST} + 2.5\, \mathrm{log}_{10}(A)
\end{equation}

\noindent where $m_{HST}$ is the total signal of the annulus, in AB magnitudes, and $A$ is the area of the annulus in arcsec$^{2}$. 

We perform background subtraction in the same manner as we do in our aperture photometry, but now we apply it to each annulus. Furthermore, we mask nearby clumps that may contaminate the surface brightness profile of a given clump, by excluding those pixels from our calculations. Finally, we are also able to derive colour profiles from the surface brightness profiles.

In Figure \ref{fig:g04-1-c2}, we show examples of the surface brightness (left panel) and colour profiles (middle and right panel) for clumps in galaxies D13-5 (clump 4), G08-5 (clump 3), G14-1 (clump 3), and G20-2 (clump 9). We chose these four examples to illustrate two typical trends we observe in the clump colour profiles.  We measure the lowest S/N and largest uncertainties in the F225W images, thus the $225-336$ colour profiles generally have larger uncertainties than the $336-467$ colour profiles. The filled coloured symbols in the middle and right panels correspond to the actual colour measurements, while the open symbols show the conversion of these colours into $A_{V}$ extinction and age. 

To map our colours to age and extinction, we first note that:

\begin{equation}
    \frac{A_{225}}{A_{V}} = 2.667089, \frac{A_{336}}{A_{V}} = 1.634535, \frac{A_{467}}{A_{V}} = 1.2229745 
\end{equation}

\noindent assuming the \citet{cardelli89} extinction law. Combining these relations, we obtain:

\begin{equation}
    A_{225} - A_{336} = C_{1} \times A_{V} = (225-336)_{o} - (225-336)_{i}
    \label{eq:c1}
\end{equation}

\begin{equation}
    A_{336} - A_{467} = C_{2} \times A_{V} = (336-467)_{o} - (336-467)_{i}
    \label{eq:c2}
\end{equation}

\noindent where the subscript ``$o$'' corresponds to observed colours and the subscript ``$i$'' corresponds to intrinsic colours, and $C_{1} = 1.032554$ and $C_{2} = 0.411561$. Then, multiplying equation \ref{eq:c1} by $C_{2}$ and equation \ref{eq:c2} by $C_{1}$, and subtracting the two equations from each other, we obtain an equation independent of A$_{V}$:

\begin{equation}
    C_{2}[(225-336)_{o} - (225-336)_{i}] - C_{1}[(336-467)_{o} - (336-467)_{i}] = 0.
    \label{eq:solve-ages}
\end{equation}

\noindent We use our observed colours from the HST observations and the colours generated by \texttt{Starburst99} with 0.1~Myr time steps to solve equation \ref{eq:solve-ages} and infer an age. We then use the results to solve for A$_{V}$ from equations \ref{eq:c1} and \ref{eq:c2}. Those are the numbers that we present on the right y-axis in both of these panels. We match to extinction and age in this way because these two parameters are not entirely independent of one another; i.e., changes in extinction have a small effect on the $336-467$ colour, while changes in age also have a small effect on the $225-336$ colour (the model grid lines are not parallel to the x$-$ and y$-$ axes). A significant fraction of the clumps in our sample show colour (age/extinction) profiles such as the ones we show for galaxies D13-5 and G08-5 in Figure \ref{fig:g04-1-c2}, where there is a strong increase in age, while fewer clumps show profiles such as the ones we show for galaxies G14-1 and G20-2, where there is no strong change in age but a strong change in extinction across the clump. This will be discussed further in \S\ref{subsubsec: tracks}.

\section{Results and Discussion} \label{sec:results}
\subsection{Galactocentric Variations of Global Clump Properties} \label{subsubsec: gradients}
\begin{figure*}
    \centering
    \includegraphics[width=\textwidth]{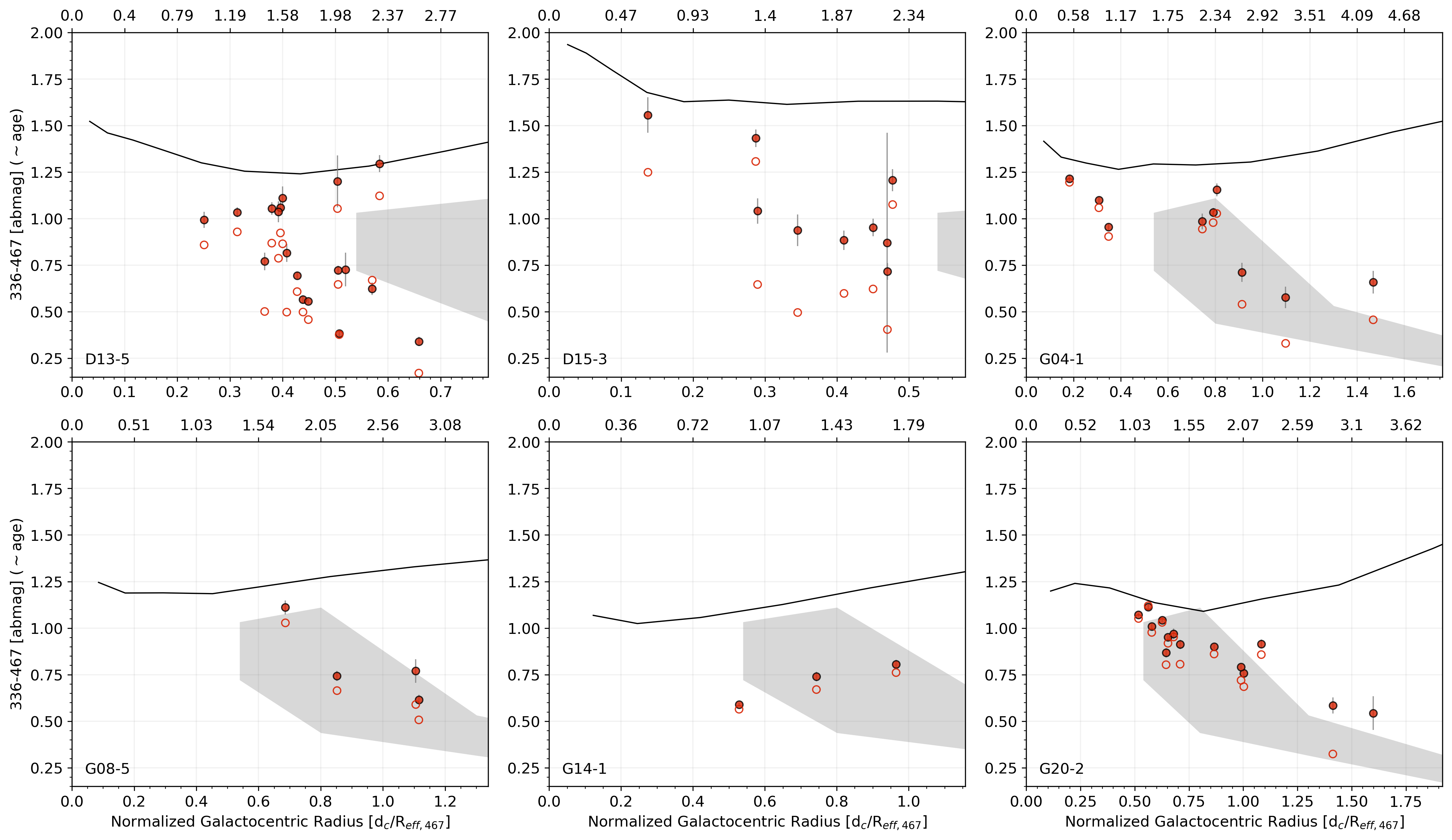}
    \caption{Clump integrated $336-467$ colour without (red filled circles) and with disc subtraction (red open circles) as a function of galactocentric radius (the upper x-axis scale gives the galactocentric radii equivalents in kpc). Photometric uncertainties appear as gray error bars. The gray-shaded regions are the $u-b$ results of \citet{guo18} at redshifts of $0.5 \leq z < 3.0$, for comparison. The $336-467$ colour, a proxy for age, is redder for clumps closer to the center of their galaxy and bluer for those on the outskirts, with the exception of galaxy G14-1 for which the opposite is true. However, the disc colours are relatively flat in comparison (black solid line). Furthermore, we see that clumps are on average bluer than the disc, and this is true whether or not we subtract the disc light. This suggests clumps are older closer to the centers of their galaxies and that this is not an effect of contamination from disc stars.}
    \label{fig:colour_radius_336}
\end{figure*}

\begin{figure*}
    \centering
    \includegraphics[width=\textwidth]{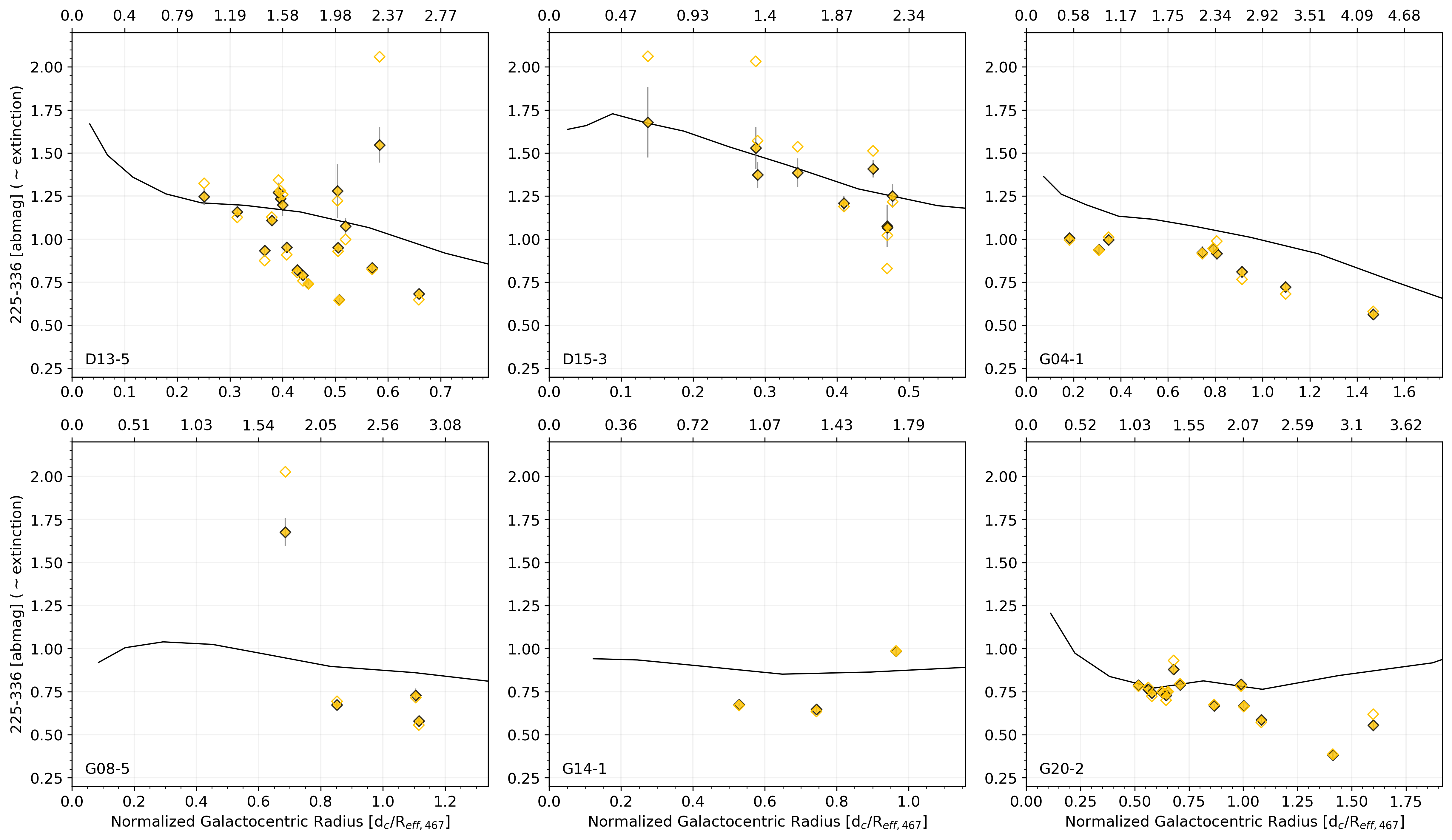}
    \caption{Same as Figure \ref{fig:colour_radius_336}, but now showing the integrated $225-336$ colour without (filled yellow diamonds) and with (empty yellow diamonds) disc subtraction. The gray solid line shows how the $225-336$ colour evolves across each galaxy when we mask clumps. The $225-336$ colours of clumps are generally flatter than the $336-467$ and track the disc colour more closely, suggesting clump extinction does not differ strongly from the disc extinction.}
    \label{fig:colour_radius_225}
\end{figure*}

We begin by first inspecting the evolution of clump colours as a function of their position within their galaxy, and compare clump colours to the disc colours. In Figure \ref{fig:colour_radius_336} and \ref{fig:colour_radius_225}, we plot for all six galaxies, the integrated $336-467$ and $225-336$ colours respectively, of each clump as a function of galactocentric radius, normalized to the galaxy F467M half-light radius (effective radius: R$_{eff,467}$). To assess the impact of disc light on the clump colours we measure, we show the non-disc subtracted (filled symbols) and disc subtracted clump colours (empty symbols). We include the photometric uncertainties (section \ref{subsubsec: ap_phot}) as gray error bars on the non-disc subtracted colour. The solid black line in each panel shows the disc light evolution. To measure the disc light, we first mask all pixels that we identify as clump pixels. We then place a series of annuli at the galaxy center, which we define as the pixel of peak emission in existing \textit{HST} F125W images (wide $J$). We finally integrate the light in each annulus, out to 4\arcsec. 

In our clump light to disc light comparison, we notice two things: (1) the $336-467$ clump colours are generally bluer than the disc light, within the uncertainties, independently of whether we subtract the disc background light from our clump measurements or not, and (2) the disc light is much flatter as a function of radius, than the clump colours \citep[these trends were also observed for $0.5 \leq z < 3.0$ galaxies by][]{guo18}. With respect to the first point, it is perhaps not surprising that the clumps exhibit bluer colours than the disc, because we select them as bright regions in the F336W (U band) images. This does however suggest that clumps are a structure separate from and embedded within the disc of the galaxy. This is further supported by our second observation: that the clump colour gradient is much steeper than the disc light suggests that clumps are structures that exist and evolve within the discs.

We see in all cases except for G14-1 that the $336-467$ colours of clumps tend to be redder closer to the galactic centers, and bluer farther in the outskirts. This trend has been previously observed in the $u-b$ colours of high$-z$ galaxies \citep[e.g.,][]{forsterschreiber11,soto17,guo18}. We reproduce here the results presented in Figure 8 of \citet{guo18} as the gray shaded area in each panel of Figure \ref{fig:colour_radius_336}, to illustrate the trends measured by these authors. The authors normalize their clump distances to the HST F160W effective radius along the galaxy major axis. At redshifts of $0.5 \leq z < 3.0$, they are unable to distinguish clumps from galaxy bulges at normalized galactocentric radii $< 0.5$, which is where we find many of our clumps. This shaded region covers the median $u-b$ values in the $0.5 < z < 1.0$ and $1.0 < z < 2.0$ redshift bins, for all three mass ranges the authors present. Such a colour gradient suggests that since the $336-467$ colour is age-sensitive, the clumps that are closer to the centers of their galaxies are older than their bluer counterparts in the outskirts. We will discuss this further in \S\ref{subsubsec: ages}.

It is important to note that our observations remain true whether we apply disc subtraction or not. In fact, we can see that when we subtract the disc light, the effects are generally the greatest on the clumps that are farther from the centers of their respective galaxies: the colours of these clumps become even bluer than the uncertainties, while the clumps closer to the center do not change so drastically. This has the effect of steepening the colour gradients we observe when we perform disc subtraction and is in disagreement with the hypothesis that colour gradients observed in galaxies are the result of contamination of clump colours by older disc stars, whose densities are larger closer to the centers of galaxies \citep{buck17,oklopcic17}. 

Other authors suggest that negative colour gradients, in which the clumps deviate from the galaxy gradient, are interpreted as supporting theories in which clumps migrate to the centers of galaxies and contribute to bulge growth. \citet{dekel09} predict migration timescales on the order of $\sim10\times$ the dynamical timescale, or about 0.5 Gyr, from their theoretical framework, while \citet{ceverino10} perform the first cosmological simulation of high-$z$ clumpy discs and show that their simulated clumps migrate to the centers of their hosts in $\sim8\times$ the dynamical timescale. From the clump ages suggested by the integrated colours in Figure \ref{fig:colour_colour} of $\sim 50-250$ Myr, and the colour gradient observed here, we find that this scenario remains plausible given these observations and our measurements.

In Figure \ref{fig:colour_radius_225}, we show the behavior of the $225-336$ colour and compare the clumps to the discs (as in Figure \ref{fig:colour_radius_336}). We see that the disc colour gradient varies from galaxy-to-galaxy: a flat profile in this colour suggests little variation in extinction across the disc of the galaxy, while a downward (upward) turn in the colour would suggest a decrease (increase) in extinction. When inspecting the clump colours, we do not find a very strong preference for clumps to appear much bluer than the discs, as we do in the $336-467$ colour. Clumps are generally either slightly bluer than the discs or scatter around the disc colours. It may not be surprising to measure bluer colours for clumps, because they are F336W-bright regions, and thus are likely to have less dust. In fact, previous works have observed this trend as well; for example \citet{wuyts13} find that the rest-UV-selected clumps in their sample of 473 CANDELS galaxies exhibit lower H$_{\alpha}$/UV ratios than the underlying disc, which they interpret as an indication that these clumps are viewed through a smaller column of obscuring material. Conversely, \citet{bassett17} find little spatial variation in extinction across a subset of DYNAMO galaxies, thus it may also not be surprising to find clumps with $225-336$ colours similar to the discs within which they reside. 

However, the gradient of $225-336$ colour with galactocentric radius seems to in general have a similar sign gradient for the clumps as the disc light, and in some cases the clumps follows the disc colour gradient closely. This is well illustrated in G04-1, where the disc colour decreases from the center, flattens, then decreases again. This same trend is seen in the clumps and is in contrast to what we observe in the $336-467$ colour, where the trends with galactocentric radius of clumps is not observed in the disc light, indicating that disc contamination is not the source of the colour gradient. 

Our observations of the $225-336$ clump colours remain true when we apply disc subtraction. The effect of disc subtraction on the $225-336$ colour is less severe than on the $336-467$ colour and shows that dust extinction is not a contributing factor to the observed colour gradient in the $336-467$ colour. In summary, clumps are similar to their host galaxy's disc in extinction, but are very different in age.

\subsection{Internal colour Gradients of Clumps} \label{subsubsec: tracks}
\begin{figure}
    \centering
    \includegraphics[width=\columnwidth]{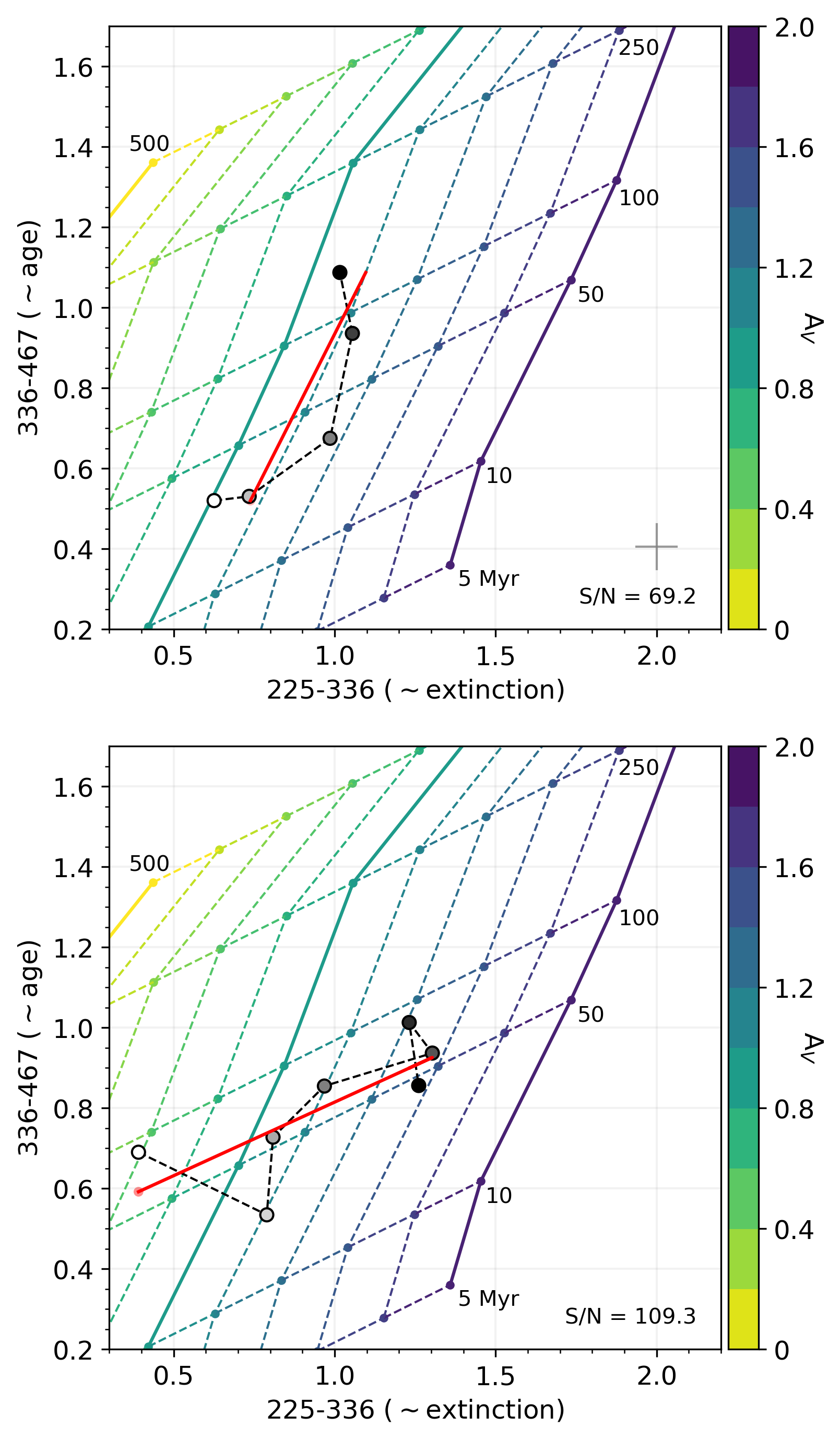}
    \caption{The same colour$-$ colour space as we show in Figure \ref{fig:colour_colour}. The data points now represent the colour track of a two clumps: DYNAMO D13-5 clump 4 (top panel) and DYNAMO G20-2 clump 9 (bottom panel), with colours measured in concentric circular annuli centered on the clump, showing their internal age and extinction structure. The shading of the data points corresponds to the distance from the center of the clump in steps of 0.04\arcsec ($\sim 60$~pc) as described in \S\ref{subsubsec:sbps}: white represents the colour at the center, black represents the colour farthest from the center. The red solid lines shows the linear fits to these colour track. The internal colour gradient of the first clump indicates that it has a constant $A_{V}$ of $\sim$\,1.2 and that the age is a few 10~Myr at the center and increases to $>100$~Myr at the clump edge, while the colour track of the second clump indicates it has a constant age of $\sim$\,50~Myr and a large change in extinction (from $A_{V}$\,$\sim$\,0.8$-$1.6). Median error bars on the colour track points are shown in the bottom right corner of the top panel.}
    \label{fig:d13-5-c4}
\end{figure}

The $0.092$\arcsec\, resolution of the F225W HST observations allows us to go beyond investigating integrated clump colour trends with galactocentric radius, and to inspect how these colours vary across the extent of each individual clump (``colour track''). At the average DYNAMO redshift of $z$\,$\sim$\,0.1, this angular resolution corresponds to $\sim$\,170~pc; thus, it is not possible to do this analysis in clumps at $z$\,$\sim$\,1. To derive the internal colour gradient of each clump, we use the surface brightness profile measurements of Section \S\ref{subsubsec:sbps} to calculate colours in concentric circular annuli, centered on each clump (essentially combining the two colour profiles in Figure \ref{fig:g04-1-c2} into a two-dimensional representation, which we call a ``colour track''). We illustrate one example of such a colour track for galaxy DYNAMO D13-5, clump 4, in Figure \ref{fig:d13-5-c4}. The axes and the coloured grid are the same as in Figure \ref{fig:colour_colour}; however, the data points now represent the colour change of one clump from its center (white point), to its edge (black point). In this particular case, we see that the colour track suggests a large change in age with fairly constant extinction across the clump. This is because as we move outward from the center of the clump, the colours track along the line of a single extinction model (in this case $A_{V}$\,$\sim$\,1.2), indicating constant extinction and varying age. For ease of visualization, we fit this colour track with a line that minimizes the distance between the points and itself. The red line in Figure \ref{fig:d13-5-c4} is this fit. We fit the colour track of all 58 clumps in our sample in this way.

\begin{figure*}
    \centering
    \includegraphics[width=\textwidth]{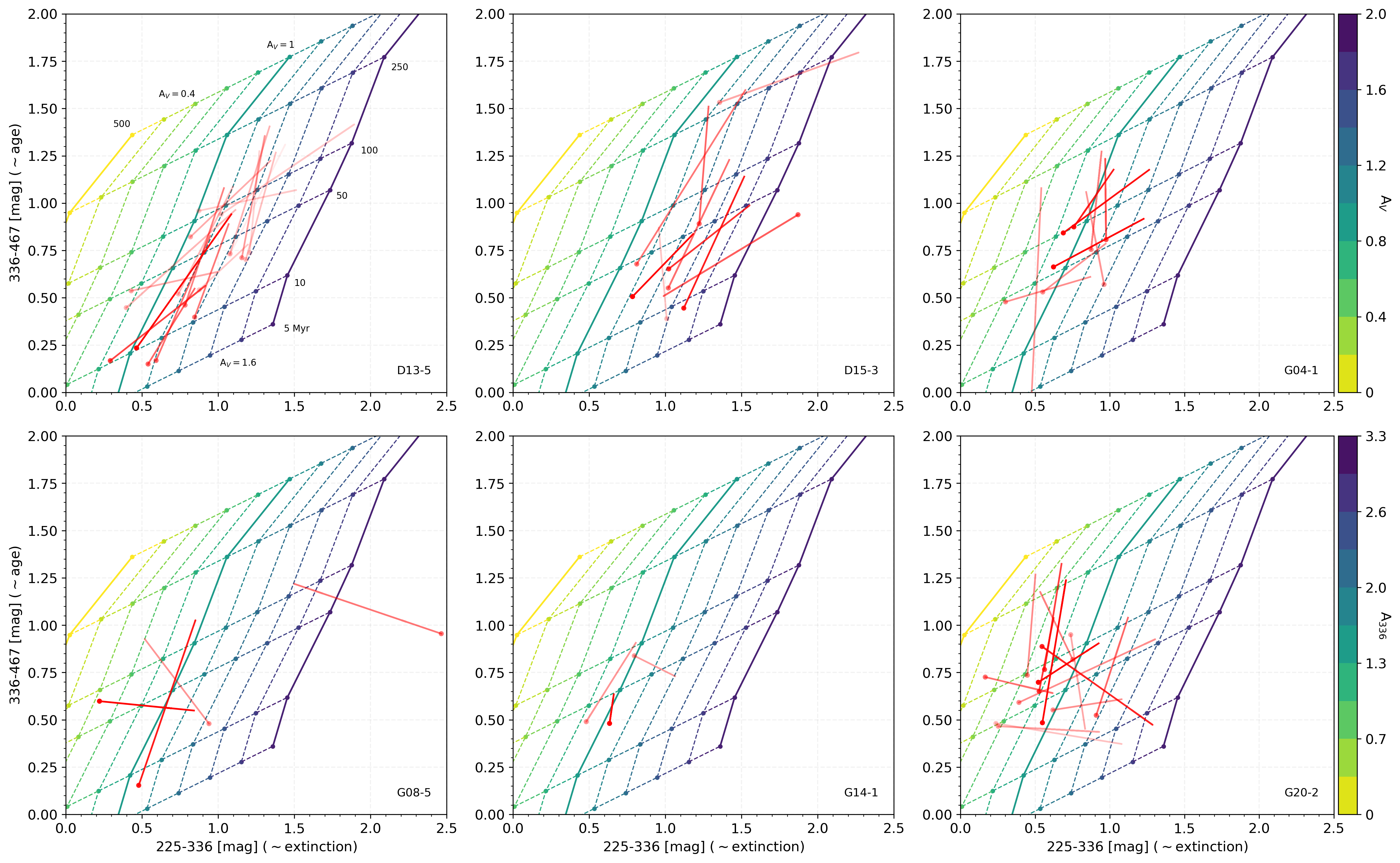}
    \caption{For each clump in our sample we plot a colour track which represents how the colour of the clump itself evolves from its center to its edge, in the $225-336$ vs $336-467$ parameter space. The coloured lines represent $V-$band extinction, as in Figure \ref{fig:colour_colour}. The red lines show the linear fit to the colours of each clump measured in concentric annuli as a function of distance from the clump center with the dot indicating the value at the center, and are shaded based on the F336W S/N (these range from $\sim$\,$10-150$). Thus, lighter shaded lines have lower S/N, while darker ones have a higher S/N and are more certain. We can see that for high S/N clumps the colours typically change primarily along the vertical axis, indicating that there is a gradient present within clumps themselves. Moreover, the majority of these clumps change in the direction of increasing age farther from their centers. A handful of the clumps show preferentially an extinction gradient, and for some galaxies (e.g., D13-5) the presence of age gradients is more marked than for others (e.g., G20-2).}
    \label{fig:colour_tracks}
\end{figure*}

In Figure \ref{fig:colour_tracks} we show the fitted lines of the colour tracks for all clumps. The axes and coloured grid are exactly the same as in Figures \ref{fig:colour_colour} and \ref{fig:d13-5-c4}. We have added an additional colour bar in the second row to show how $A_{V}$ translates into $A_{336}$ when we assume the extinction law of \citet{cardelli89}. The solid red shaded lines show the fitted colour track of each clump in our sample. Thus, they represent the general change of the $225-336$ and $336-467$ colours across the extent of each clump. We shade the lines to represent the clump S/N in the F336W images: darker (lighter) shaded lines have higher (lower) S/N. The single data points found at one end of each colour track correspond to the central region of each clump, thus indicating the direction in which the distance from the clump center increases. The length of each colour track is such that the end point corresponds to the colours of the outermost annulus of each clump, defined as the ``edge" of the clump as described in Section \ref{subsec: clumpd_ID} ($r_{clump}$).

We identify from this collection of colour tracks three scenarios (in increasing order of occurrence frequency): (1) clumps with colour tracks that suggest little to no change in age and large changes in extinction across the clump; these are indicated by shallow positive (increasing extinction) or shallow negative (decreasing extinction) slopes (i.e., colour tracks that cut across many lines of $A_{V}$ but follow a single age line) -- see galaxy D13-5, D15-3, G04-1, and G20-2 for examples, (2) clumps whose colour tracks indicate decreasing age with increasing distance from the clump center; these are indicated by steep negative slopes -- see galaxy G20-2 for examples, and finally (3) clumps whose colours show large increases in age from the clump center with $\sim$\,constant extinction (i.e., these tracks cut across many ages lines but no, or few, lines of $A_{V}$) in some cases, and with large increases (decreases) in extinction; these are indicated by steep positive (negative) slopes (i.e., these tracks cut across many lines of $A_{V}$ and many age lines) in other cases. Of the 58 clumps in our sample, the distribution of clumps per scenario is: (1) 7, (2) 10, and (3) 40. Thus, $\sim$\,70\% of the clumps in our sample show colour tracks with increasing ages away from the centers.

To verify that our PSF matching is not falsely producing these colour gradients, we create mock clumps to test our PSF matching procedure. We model clumps with a symmetric 2D Gaussian profile, with sizes matching those of our observed clumps. We convolve each of these to the PSF profile of each HST filter, to simulate observations. We measure the internal clump colour gradients of these simulated clumps, then we perform our PSF-matching and measure the internal clump colours once more. We find that in cases where we create clumps with zero colour gradients, we recover no gradient, and in cases where we create clumps with a given gradient, we are able to recover it as well. Thus, we conclude that our PSF-matching procedure is not producing artificial gradients.

Clumps in the first category, whose colours suggest constant ages and large increases in extinction may correspond to ``pseudo-clumps''. This may be an indicator that even though we identify these objects as clumps, they may be holes in extinction that create a clumpy light distribution. This was observed in the NIHAO simulation by \citet{buck17}, where non-clumpy galaxies showed clumpy $U-$band light distributions due to clumpy gas distributions and clumpy star formation events. However, if all clumps that we observe in our DYNAMO sample were the result of this, then we would expect all of their colour tracks to fall in this first category, which is not what we find.

Nearly three-quarters (40) of the clumps we select in these six DYNAMO galaxies lie in the third category: clumps with large increases in age. This trend is particularly well illustrated in galaxies D13-5, and D15-3. This suggests that the evolution of these two colours is driven predominantly by age: the clumps are younger at the centers and older toward the edges, and extinction is not a large contributor to the clump colour evolution. This observation suggests that these clumps are an entity separate from the galaxy disc, with an inner structure whereby star formation is predominantly found in the central regions. These results are interesting because they indicate that clumps observed in high$-z$ galaxies and local analogues likely consist of a combination of ``real'' physical structures, and ``pseudo-clumps'' that are a result of clumpy light distributions. The ability to distinguish these would likely be important for simulations of high$-z$ systems, and the colour tracks of clumps could potentially be used as an additional criterion for selecting and distinguishing real clumps in galaxies.

Another possibility is that these clumps are perhaps composed of smaller star forming clusters that follow the cluster pair separation-age difference relation \citep{efremov98,marcos09,grasha17}. \citet{grasha17} show that for eight local galaxies, the age difference between pairs of star forming clusters increases with their separation to the power of $\sim 0.25 - 0.6$. The authors also find that the maximal scales over which cluster pair separation and age difference are correlated, range from $\sim 200$ pc $- 1$ kpc. We are unable to resolve any potential smaller clusters within each clump and measure the cluster pair separations-age difference relation (measuring pixel-to-pixel differences is not feasible as the photometric uncertainties are too large to yield meaningful results). It is nonetheless worth noting that these clumps could break up into even smaller structures at higher resolution, which we cannot ascertain with currently available data. However, high-resolution simulations of isolated clumpy, high-$z$ discs do show substructures within giant clumps \citep[see e.g.,][]{ceverino12}.

\subsection{Galactocentric Variations of Resolved Clump Properties} \label{subsubsec: ages}
\begin{figure*}
    \centering
    \includegraphics[width=\textwidth]{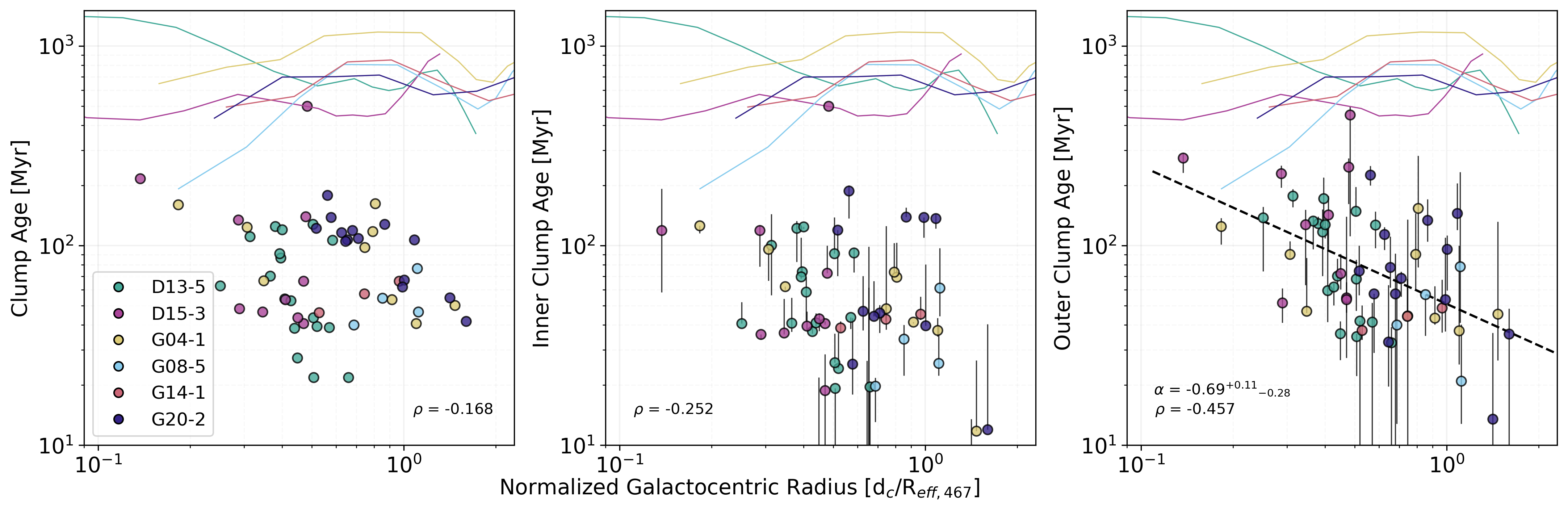}
    \caption{Clump ages estimated for the whole clump region from the integrated light within $r_{clump}$ (left), the inner 25\% regions (middle), and the outer 75\% (right), vs. galactocentric radius, compared to the estimated disc ages (solid coloured lines). Error bars show the spread in ages when shifting the inner and outer clump boundaries by 10\%. The strongest correlation is between the outer clump regions and galactocentric radius, which we quantify with the Spearman Rank-Order correlation coefficient, $\rho$, where we see clump age increasing with decreasing distance. However, the disc profiles are flat in comparison (with the exception of galaxy G08-5), suggesting that the clump age--distance relation is not due to an underlying age trend within the discs themselves. We fit a power law to the outer clump age--distance relation (black dashed line in the right panel) and report the slope of this fit ($\alpha$). The outlying point in galaxy D15-3, which appears much older than other clumps, is measured as an upper limit in the F225W image.}
    \label{fig:age_radius}
\end{figure*}

Based on the internal colour gradient observation that the majority of clumps appear older in their outskirts and younger at their centers, we compare the ``inner'' and ``outer'' clump ages as a function of galactocentric radius, shown in Figure \ref{fig:age_radius}. Here, we plot the estimated ages of clumps as a function of the clump galactocentric radius, normalized to the effective F467M radius, R$_{eff,f467m}$, in three different ways. We first consider each clump as a whole (left panel), then we split each clump at the radius corresponding to $0.5 \times r_{clump,467}$, to define an ``inner'' (middle panel) and an ``outer'' (right panel) clump region. These regions, thus, cover the inner 25\% and outer 75\% of the clump area respectively. In the last panel we also show a fit to the relationship between outer clump age and normalized galactocentric radius using a linear relation in log-log space, which corresponds to a power law in linear space. 

For the three panels in Figure \ref{fig:age_radius}, we measure the clump light in (1) a circular aperture whose radius is $r_{clump}$, (2) in a circular aperture whose radius is 0.5 $\times$ $r_{clump}$, and (3) in an annulus whose inner radius is 0.5 $\times$ $r_{clump}$ and whose outer radius is $r_{clump}$ (we check and ensure that the sum of the inner and outer regions is equal to the clump light we measure in the circular aperture of size $r_{clump}$). Performing this measurement in all three filters, we then calculate both the $225-336$ and the $336-467$ colours of these regions, and simultaneously match them to the \texttt{Starburst99} models, as described in Section \ref{subsubsec: ap_phot} and done in Figures \ref{fig:colour_radius_336} and \ref{fig:colour_radius_225}. The clumps are separated according to which DYNAMO galaxy they reside in, as indicated by the legend. The error bars on the data points indicate how the spread in estimated ages changes when we shift the boundary between the inner and outer clump by 10\%. The downward directed error bars are generally larger than the upper error bars because shrinking the inner clump boundary significantly increases the photometric uncertainties and causes clumps to appear much bluer (as a result of the intrinsic colour gradient). 

To compare the clump ages to the discs, we include the solid coloured lines which show the disc age distributions derived from the disc colour profiles, and are the same in all panels. As we show in Figure \ref{fig:colour_radius_336}, the discs are redder in $336-467$ and thus considerably older than the vast majority of the clumps. We include the non-parametric Spearman rank correlation coefficient, $\rho$, in each panel, showing that the most significant correlation is between outer clump age and galactocentric distance. Finally, we also include in the right panel the fitted slope of the outer clump age--distance relation: $\alpha=-0.69^{+0.11}_{-0.28}$, where the uncertainties correspond to 1$\sigma$ determined from a bootstrap analysis. We compute the mean age and 1$\sigma$ standard deviation of each panel to further quantify the spread in the data: (left) 90.5\,$\pm$\,69.1~Myr, (middle) 66.2\,$\pm$\,70.7~Myr, (right) 96.8\,$\pm$\,75.3~Myr.

From the left panel of Figure \ref{fig:age_radius}, we can see that clumps are generally estimated to be $\sim$\,a few 100~ Myr younger than the galaxy discs, and the ``oldest" clumps we see are $\sim$\,250~Myr old. We also do not see any ``old'' clumps in the outskirts of the galaxies, and we do not see any very ``young'' clumps close to the centers. We note that we generally observe fewer clumps at the smallest galactocentric radii; however, this may be expected as the infall of clumps closer to the centers of galaxies would take place faster than for clumps farther out. Thus it would be more likely to observe clumps at large galactocentric radii, than at smaller ones. Alternatively, it may also be a geometric effect where we would expect to find fewer clumps in a galaxy if we decrease its size (radius), and therefore its area.

When we break up each clump into an inner and outer region, we see that in general the ages of the inner clumps are younger than the outer clumps by a factor of 2 to 4. Furthermore, outer clump regions in clumps closer to the centers of galaxies tend to be older, while the inner regions of clumps have similar ages regardless of where in the disc the clump is located. In other words, the inner clump ages do not show a trend with galactocentric radius, while the outer clump regions present a much clearer trend. We can see this well illustrated when comparing the two measures of the age of a clump for a given galaxy. For example in D13-5, the inner clump ages follow a relatively flat distribution with a lot of scatter, while the outer clump ages increase toward smaller galactocentric radii. If clumps keep forming stars in their cores as they migrate through the disc, then young stellar populations will be maintained in these regions and skew the light-weighted ages to younger values. In contrast, the outer regions of clumps show a stronger preference to be older when located closer to the centers of galaxies, and younger at the outskirts. \citet{forsterschreiber11} find a steep relation between clump age and distance in their study of clumps in six $z \sim 2$ galaxies (see their Figure 8, bottom panel), with a power-law slope of $-2.06 \pm 0.63$ \citep[slope measured by][]{genel12}. In contrast, \citet{guo18} find an increasingly steeper clump age--distance relation with decreasing redshift (see their Figure 9): at $z \sim 2-3$ this relation is flat, while at $z \sim 0.5-1$ the slope of the relation becomes $\sim -1.5$. Similarly, \citet{zanella19} find a relatively flat clump age--distance relation for their sample of 53 star forming galaxies at $z \sim 1-3$ (see their Figure 8). However they list as possible explanations for the difference between lower- and higher-redshift results as: (1) signatures of migration might not yet be visible at $z \sim 2$, (2) the improved spatial resolution at lower-redshift may improve the detection of clumps at small galactocentric radii, and (3) measurement uncertainties may ``hide'' underlying age gradients. 

\begin{figure}
    \centering
    \includegraphics[width=\columnwidth]{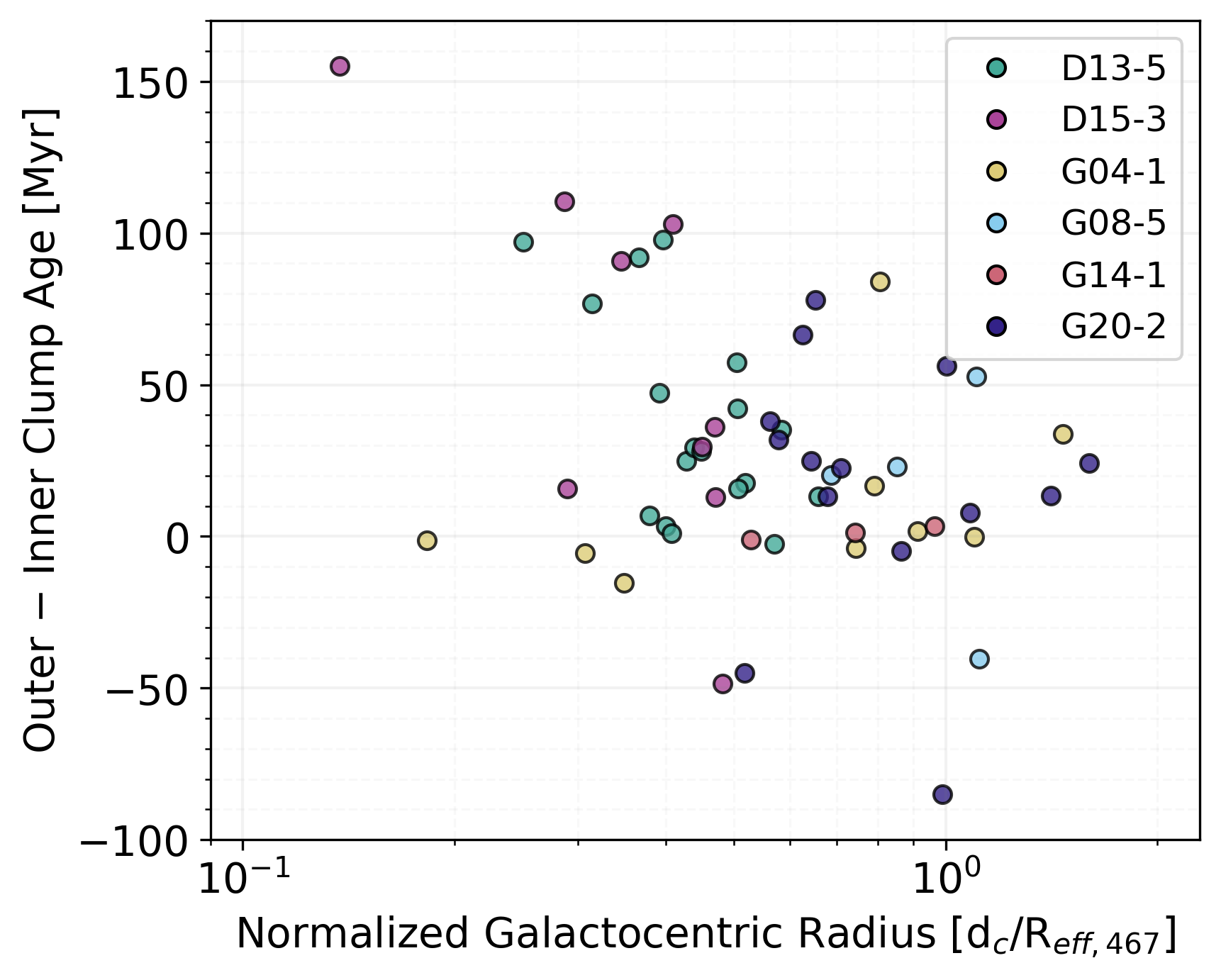}
    \caption{The age difference between the outer and inner clump region as a function of normalized galactocentric radius. Negative age differences correspond to clumps whose colour tracks indicate ``older'' inner clump regions compared to the outer clump regions. In general, the age difference increases with decreasing galactocentric distance, with the innermost clumps showing age differences as large as $\sim$\,100$-$150~Myr.}
    \label{fig:age_diff}
\end{figure}

Finally, in Figure \ref{fig:age_diff}, we plot the age difference between the inner and outer clump regions (middle and right panels of Figure \ref{fig:age_radius}) as a function of normalized galactocentric distance. Clumps with negative age differences are those with colour tracks that show decreasing ages with increasing distance from the clump center. For the remaining clumps, we generally observe the smallest age differences for clumps at the largest galactocentric distances, and the largest age differences for the more centrally located clumps. For these, we observe age differences as large as 150~Myr. \citet{dekel09} predict that if radiative feedback successfully suppresses star formation in clumps, then the spread in their stellar ages should not exceed $\sim$\,50$-$100~Myr. 

\subsection{Comparison to Hydrodynamic Models}
While there is a mixture of results at higher-redshifts, our picture is consistent with the findings of high$-z$ simulations such as those presented in \citet{ceverino10,ceverino12,bournaud14,mandelker17}. \citet{ceverino10} present results from the first cosmological simulations of three high$-z$, clumpy galaxies that produce disc fragmentation from gas stream driven instabilities. This results in discs with morphologies that are very similar to observed high$-z$ systems, including clumpy star forming rings such as we see in DYNAMO D13-5 (our Figure \ref{fig:obs1}; see their Figure 6). The resulting clumps migrate and coalesce with the bulge on timescales of $\sim$\,250~Myr. This timescale is consistent with the outer clump ages of the most centrally located clumps in our sample (see Figure \ref{fig:age_radius}). \citet{ceverino12} build on this by including two additional zoom-in cosmological simulations at $z = 2-3$ to study internal support against collapse for 77 clumps. This works includes three clumps analysed at 2~pc resolution, where they find that they fragment into dense subclumps and that resolving these is important for understanding the internal structure of clumps. 

\citet{bournaud14} identify a population of massive ($\sim 10^{8}$ M$_{\odot}$) clumps that remain long-lived in the simulations, meaning that they can be tracked in the simulations for $200-500$ Myr, and up to 700 Myr in some cases, which migrate to the centers of their galaxies. These clumps maintain relatively constant masses and star formation rates ($1-2$ M$_{\odot}$\,yr$^{-1}$), despite outflows ($1-2$ M$_{\odot}$\,yr$^{-1}$ across a 1 kpc$^{2}$ section) and the dynamical loss of older stars ($0.2-2$ M$_{\odot}$\,yr$^{-1}$), thanks to re-accretion of gas from the surrounding gas-rich discs ($2-15$ M$_{\odot}$\,yr$^{-1}$). The observational consequence of this is that the apparent ages of clumps are dominated by the light of young stars, and therefore saturate at around 200 Myr, thus clumps can be up to twice as old as the median age of the stars within them. This is consistent with the lack of clumps much older than $\sim$ 200 Myr that we observe, while the tendency of older clumps to be preferentially located towards the centers of galaxies supports the clump migration scenario. Furthermore, if the re-accreted gas falls to the centers of clumps and forms stars there, then this would create clumps with young centers and older outskirts, consistent with the clump age substructures we observe in our sample.

Similar results are found by \citet{mandelker17}, who also find that the stellar ages of clumps are less than the true lifetime of clumps due to ongoing star formation. In this work, the authors study the properties of high$-z$ galaxies in zoom-in hydro-cosmological simulations with and without radiation pressure feedback. The authors find that even in the case of radiation pressure feedback, clumps more massive than 10$^{8.2}$~M$_{\odot}$ are able to survive in the disc and migrate to the centers of galaxies, while less massive clumps are considered short-lived \citep[as long as their ages are $\lesssim 20\,\times$ the clump free-fall timescale; see][\S\,5.1.2 for details]{mandelker17}. The authors present a comparison of the ages of short-lived clumps, long-lived clumps, and \textit{ex situ} clumps (which are clumps that join the disc after a merger event), as a function of galactocentric radius (their Figure 15, top left panel). 

\begin{figure}
    \centering
    \includegraphics[width=\columnwidth]{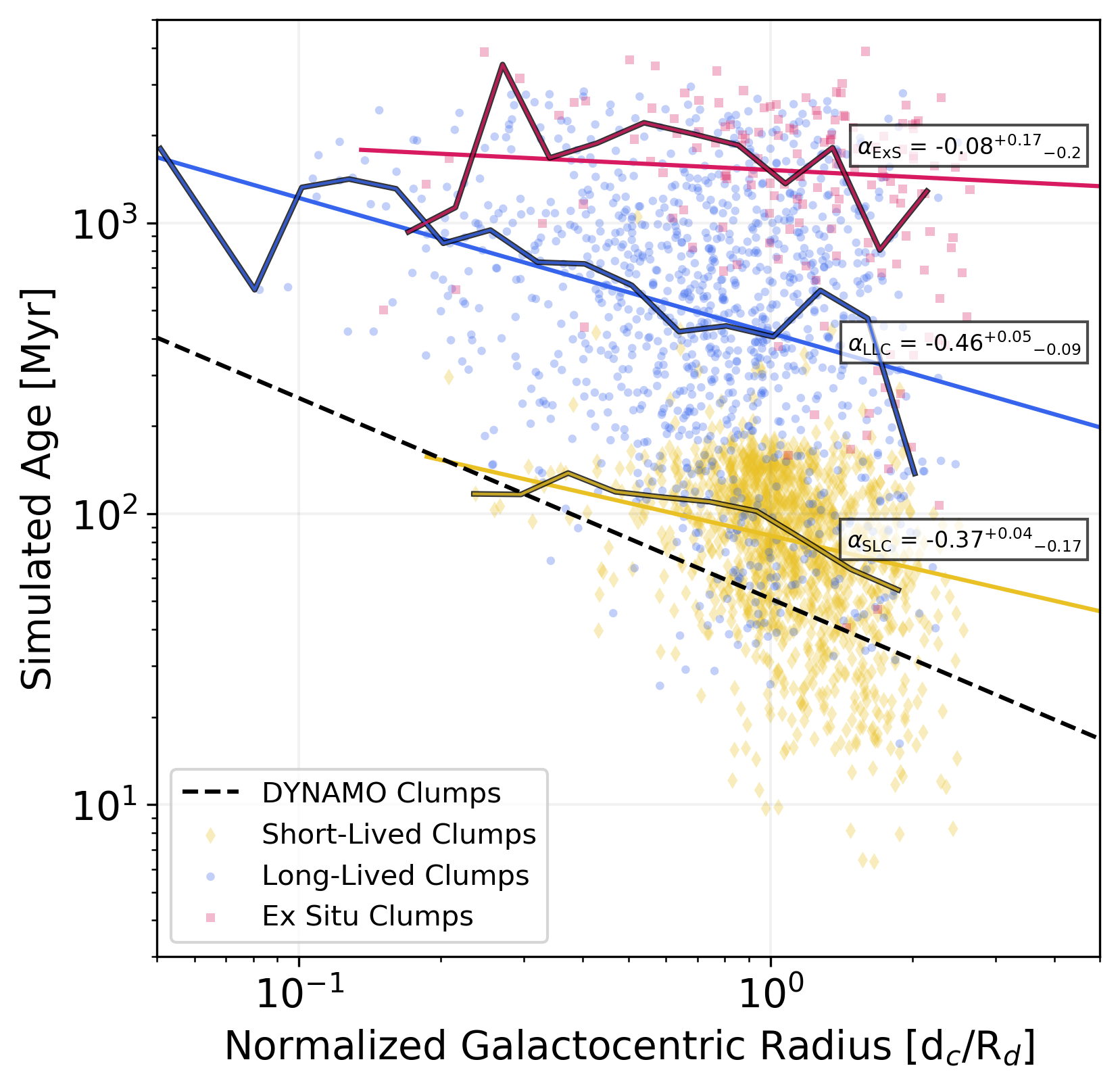}
    \caption{Comparison of the short-lived (yellow diamonds), long-lived (blue circles), and \textit{ex situ} (pink squares) clumps of \citet{mandelker17}. The solid coloured lines with black outlines correspond to the 50$^{th}$ percentile in bins of log normalized galactocentric radius, while the coloured lines with no black outline represent the power law fits to the 50$^{th}$ percentiles. The power law slopes with uncertainties determined through a bootstrap analysis are included in the figure. For reference, the black dashed line is the best power law fit to the outer clump ages which we show in Figure \ref{fig:age_radius}.}
    \label{fig:mandelker}
\end{figure}

We reproduce these results in Figure \ref{fig:mandelker}, where the coloured data points represent the long-lived clumps (blue circles), short-lived clumps (yellow diamonds), and \textit{ex situ} clumps (pink squares), while the solid coloured lines with black outlines represent the 50$^{th}$ percentile of the respective three clump populations, binned in the log of the normalized galactocentric distance. We can see from the data points in Figure \ref{fig:mandelker} that the scatter is large for all three populations of clumps. Two factors make a direct comparison difficult: (1) the galactocentric distances in \citet{mandelker17} are normalized to $R_{d}$, the radius that contains 85\% of the cold mass within a cylinder of a certain size \citep[see][for a more detailed description]{mandelker17}, and (2) the ages of simulated clumps are measured as the mass-weighted mean stellar age. Neither the cold mass distribution nor the mass-weighted age can be easily measured in observations: the lack of a good handle on the first introduces an uncertainty in the scaling of the x-axis, while the light-weighted stellar ages we observe will be systematically younger than mass-weighted ages. As a result, a direct comparison of the DYNAMO clump age estimates to the simulated clump ages is not meaningful. However, we can compare the slope of the power law fit to the DYNAMO outer clump ages as a function of galactocentric distance, to the slopes of the power law fits of the three clump populations of \citet{mandelker17}. To determine the slopes, we fit a power law to the 50$^{th}$ percentile lines; these fits are represented by the coloured lines (no black outline). Finally, we perform a bootstrap analysis to determine the uncertainties of these three fits. The three slopes along with their 1$\sigma$ uncertainties are shown on the figure. They are: $\alpha_{\mathrm{SLC}} = -0.37^{+0.04}_{-0.17}$, $\alpha_{\mathrm{LLC}} = -0.46^{+0.05}_{-0.09}$, and $\alpha_{\mathrm{ExS}} = -0.08^{+0.17}_{-0.19}$. In comparison, the slope we find for the DYNAMO outer clump ages is: $\alpha = -0.63^{+0.15}_{-0.13}$. 

From this analysis, we are able to conclude that the properties of the DYNAMO clumps are distinct from the \textit{ex situ} clumps, and thus that the clumps we observe are likely not \textit{ex situ} clumps. This conclusion is also reported by \citet{huertas-company20}. The authors present a new neural networks based method for selecting clumps, which they apply to 9,000 star forming galaxies in CANDELS in the redshift range $z \sim 1-3$. Their method yields $\sim$3,000 clumps in $\sim$1,500 galaxies. They derive clump properties, including stellar mass, from spectral energy distribution fitting of seven photometric bands, and then construct a clump stellar mass function. The authors apply the same procedure to 35 galaxies from the VELA simulations \citep{ceverino14} and find that, under the same conditions, the cluster stellar mass functions agree well, suggesting an \textit{in situ} origin for clumps in CANDELS. 

The slope of the age--distance relations for DYNAMO clumps may be as steep as $-0.76$ or as shallow as $-0.48$, which puts it within the upper bound of the long-lived clumps slope (ranges from $-0.41$ to $-0.55$). However, the slope of the short-lived clumps ranges from $-0.33$ to $-0.54$, thus it is not entirely clear if the outer clump regions of DYNAMO clumps are more similar to short-lived or long-lived clumps in the high$-z$ simulations of \citet{mandelker17}. This is a result of the scatter at large galactocentric radii, which drives the large uncertainty toward steeper slopes for both the short-lived and long-lived clumps. We note that in their high resolution cosmological simulations of $z \sim 2$ galaxies, \citet{genel12} find a power law slope of -0.57\,$\pm$\,0.14 for the relation between clump age and distance, which is in good agreement with our result. However, the authors explain that their age gradient is due to the dominance of ``background'' stars which are a part of the galaxy discs and not actually gravitationally associated with the clumps. Then, since their galaxy discs themselves have a negative age gradient (older disc stars near the centers of galaxies), the clumps that are at found at smaller galactocentric radii also appear older, despite no evidence of clump migration. Thus they conclude that the clump-distance relation is a consequence of inside-out disc formation. In contrast, the colours we measure for the DYNAMO galaxy discs do not imply such strong negative age gradients, as was also observed in \citet{ceverino12} who found a slope of $-1.2$ for their simulated clump age--distance relation and a slope of only $-0.3$ for the simulated discs; thus we conclude that the age gradients we observe in the clumps are not driven by the stellar distributions of the discs.

\begin{figure}
    \centering
    \includegraphics[width=\columnwidth]{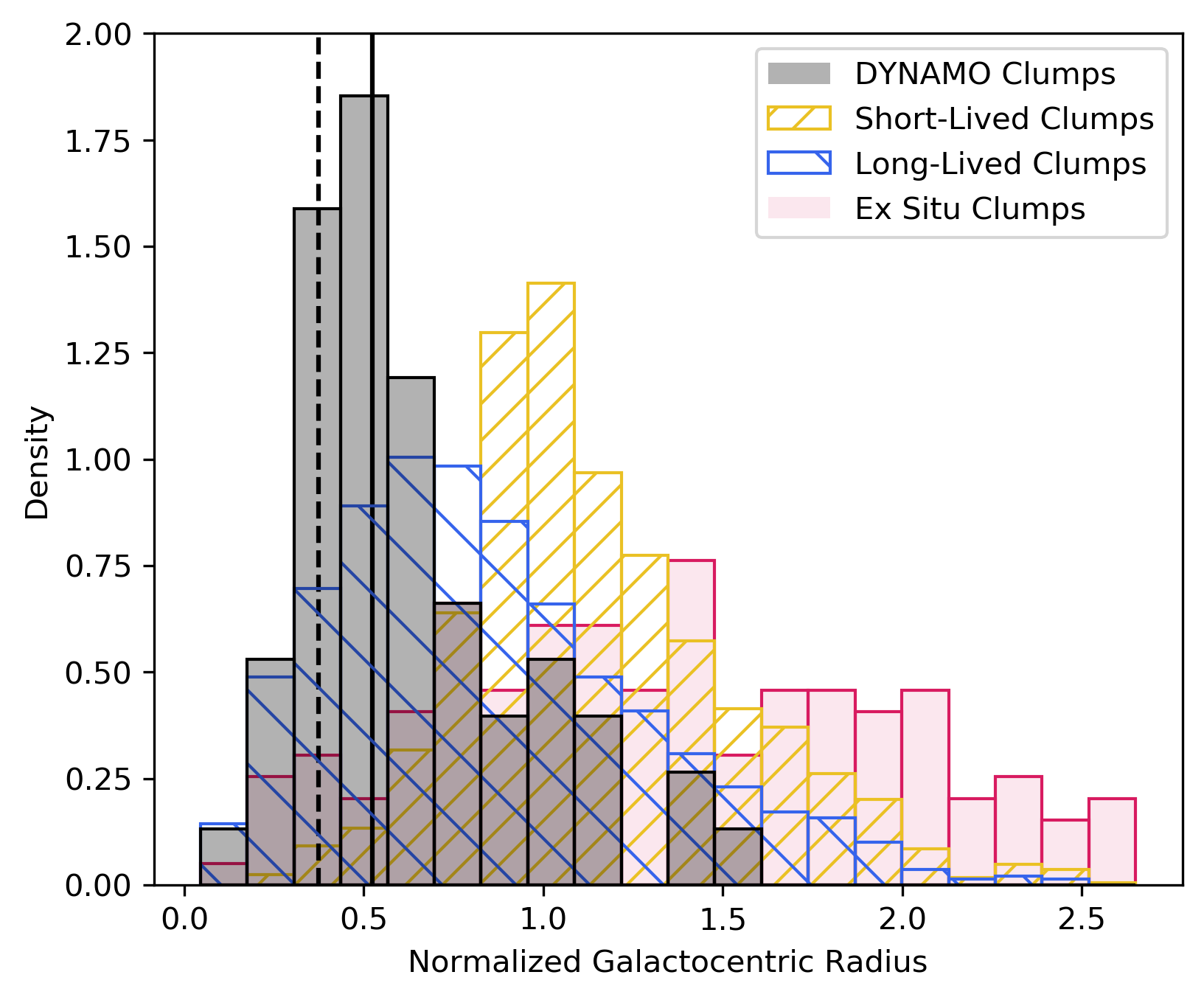}
    \caption{Comparison of the distribution of DYNAMO (grey), short-lived (yellow), long-lived clumps (blue), and \textit{ex situ} clumps in terms of normalized galactocentric radius. The DYNAMO clumps are clearly distinct from the short-lived clumps and cluster at radii $< 1$, as do the long-lived clumps. The median distance of DYNAMO clumps (vertical black solid line) is $\sim$\,0.5 and is $\sim$\,1.1 for the short-lived clumps. The radius normalization of the simulated clumps corresponds to the radius that contains 85\% of the cold mass, while we normalize to the 50\% light-radius measured in the F467M images. If we choose a larger radius definition as our normalization, the median distance of DYNAMO clumps becomes $\sim$\,0.5 (vertical dashed black line).}
    \label{fig:distances}
\end{figure}

As a final point, we note that the simulated short-lived clumps are not common at very small radii. Thus, despite the uncertainty introduced by the different radius normalization between our work and that of \citet{mandelker17}, we draw a comparison between the distribution of DYNAMO clumps in galactocentric radius, to the distribution of the short-lived, long-lived, and \textit{ex situ} clumps in Figure \ref{fig:distances}. We can see that the short-lived clumps cluster at larger galactocentric radii than do the long-lived and DYNAMO clumps. Thus the DYNAMO clumps and short-lived clumps appear as two distinct populations in this figure. In fact, 50\% of the simulated short-lived clumps reside within a normalized galactocentric radius of $\sim$\,1.1, while 50\% of the DYNAMO clumps are found within a distance of $\sim$\,0.5; this implies that a large fraction of our clumps are not short-lived.  We use the 50\% light-radius in the F467M filter as our normalization for each galaxy; however, if we instead normalize to the 85\% light-radius in the F467M filter the peak of the distribution shifts to slightly smaller distances. The vertical black solid line indicates the median galactocentric radius of DYNAMO clumps when normalized to the 50\% light-radius, while the vertical, dashed black line indicates the shifted peak when we normalize to the 85\% light-radius. This shift represents a change of a factor of 0.15 dex. Thus, we tentatively conclude from these comparisons that the radial distribution of the DYNAMO clumps is more similar to the long-lived clumps than the short-lived clumps identified in the high$-z$ simulations of \citet{mandelker17}.

\subsection{Toward a General Picture of Star Formation in Clumps}
Putting the results of our HST photometry and colour measurements together, we have a picture which is consistent with clump ages ranging from 10$-$250~Myr, with older appearing clumps preferentially located closer to the centers of their galaxies. We find that more than half of the clumps in our sample have an internal age gradient, such that stellar population age increases toward the clump edges. We interpret this as a scenario where continuous star formation in the centers of clumps replenishes the populations of massive young stars as the clumps age and move through their galaxy disc. This is consistent with the detailed picture of long-lived clumps from simulations \citep{ceverino10,ceverino12,bournaud14,mandelker17}.

White et al. (submitted) find with AO-enabled K-band imaging from Keck/NIRC2 that the stellar masses of clumps in DYNAMO G04-1, included in this sample, are preferenced toward low masses. The clumps with $M_{star}>10^8$~M$_{\odot}$ are located in the center of the galaxy (after accounting for disc subtraction). This is consistent with our picture of long-lived stellar clumps that grow as they migrate to the galaxy center.

The range of ages we find, the trend of increasing age with decreasing galactocentric radius, and the young clump centers are all consistent with high$-z$ simulations that show the existence of long-lived clumps that are able to migrate to the centers of their galaxies. The distribution of our clumps in terms of normalized galactocentric radius is consistent with the long-lived clumps of \citet{mandelker17}, which further supports this scenario. 

Our data are not consistent with predictions that clumps are only a result of extinction gradients \citep{buck17}, as we observe that more than half of our clumps have colour tracks that indicate large variations in age with little variation in extinction. Furthermore, we propose that internal clump colour gradients may be potentially used as a discriminator between clumps in mass and clumpy light distributions, and suggest that observed clumps in high$-z$ samples may consist of a combination of these.

These results point to a depiction of clumps, not as monolithic structures, but as complex objects with substructures that require further study. A comparison of high resolution clump properties and their substructure with high resolution simulations would be very interesting in this regard. Several simulations of gas-rich, clumpy star forming high-redshift galaxies that study the effect of resolution on observed clump properties find significant substructures within clumps \citep{ceverino12,behrendt16,behrendt19,faure21}. The giant clumps that are ubiquitous at $z \sim 1-3$ appear in these simulations to be collections of smaller clumps with masses of 10$^{7-8}$~M$_{\odot}$ and sizes on the order of 100~pc. When matched to the spatial resolution of high-redshift observations, these smaller clumps appear to coalesce into these giant clumps with very well-matched properties including size, mass, and high intrinsic velocity dispersions.

\section{Conclusions} \label{sec:con}
In this work we use HST observations of six DYNAMO galaxies taken in the F225W, F336W, and F467M filters. We perform aperture photometry and surface brightness photometry on these data, and use the results to measure the integrated and clump$-$resolved $225-336$ and $336-467$ colours of clumps. These colours are sensitive to extinction and age respectively: while the two colours are not independent of each other, changes in extinction result primarily in changes in the $225-336$ colour, and changes in age are reflected primarily in changes in the $336-467$ colour. Our main findings are: \\

\noindent 1. The integrated $336-467$ colours of clumps become increasingly redder with decreasing galactocentric radius. This indicates that clumps closer to the centers of galaxies are older than ones farther out. Furthermore, the underlying colour distributions of the galaxy discs are comparatively much flatter than the trends we observe in the clump colours. This implies that contamination from disc light is not the source of the trend we find. This is consistent with the findings of \citet{guo18}. \\

\noindent 2. Comparing the integrated clump colours to outputs from \texttt{Starburst99} colours, we find that their colours are consistent with ages between 10 and 250~Myr, and $A_{V}$ extinction between 0.6 and 2.0. The lack of significantly older clumps is consistent with the simulations of clumps in high$-z$ galaxies from \citet{bournaud14}, who shows that the ages of clumps saturate at $\sim$\,250~Myr because the clumps simultaneously experience ongoing star formation and the loss of older stars through dissipation. \\

\noindent 3. We analyze the internally resolved colours of clumps and find that for the majority of them the $336-467$ colour is progressively redder toward their edges; based on our stellar population modeling, this is consistent with an age gradient such that age increases from their centers to their edges. Clump centers tend to be fairly young, and their age does not show a measurable trend with galactocentric distance. The outer regions of clumps, on the other hand, do show a significant trend with galactocentric distance and a tighter correlation than the clump ages from the integrated light. This is likely caused by ongoing star formation in clump centers throughout their lifetime, so we take the age of the stellar population in the outer clumps to be more indicative of the actual age of the clump. \\

\noindent 4. We compare the outer ages of DYNAMO clumps to the \citet{mandelker17} high$-z$ simulations, by analyzing the slopes of the power law fits of the \textit{ex situ}, short-lived, and long-lived clumps. We determine from this comparison that the DYNAMO clumps are distinct from the \textit{ex situ} clumps. When comparing the distribution of DYNAMO clumps and simulated clumps in terms of galactocentric radius, we find a clear distinction between the short-lived clumps and DYNAMO clumps. The majority of short-lived clumps are found at larger normalized galactocentric radii, while the DYNAMO clumps preferentially cluster a smaller distances, much like the long-lived clumps.

\section*{Acknowledgements}

The authors thank the referee, Dr. Daniel Ceverino, for helping to improve this work through thoughtful comments and feedback. The authors thank Nir Mandelker for providing the simulation data presented in Figures \ref{fig:mandelker} and \ref{fig:distances}. The authors thank the Galaxy Formation group members at the Flatiron CCA for useful and stimulating discussions. L.L. would like to thank Petr Pokorn\'{y} for the many productive discussions and advice. L.L. and A.D.B. acknowledge partial support from the NASA HST-GO-15069.002-A grant. D.B.F. and K.G. acknowledge support from the Australian Research Council (ARC) Future Fellowship FT170100376, and by the Australian Research Council Centre of Excellence for All Sky Astrophysics in 3 Dimensions (ASTRO 3D), through project number CE170100013. K.G., R.G.A., and D.B.F. acknowledge past support from ARC Discovery Project DP160102235. D.B.F. acknowledges support from Australian Research Council (ARC)  Future Fellowship FT170100376. K.G. acknowledges support from ARC Laureate Fellowship FL180100060. D.O. is a recipient of an Australian Research Council Future Fellowship (FT190100083) funded by the Australian Government. This research is based on observations made with the NASA/ESA Hubble Space Telescope obtained from the Space Telescope Science Institute, which is operated by the Association of Universities for Research in Astronomy, Inc., under NASA contract NAS 5–26555. These observations are associated with program 15069. This research has made use of the NASA/IPAC Extragalactic Database (NED) which is operated by the Jet Propulsion Laboratory, California Institute of Technology, under contract with the National Aeronautics and Space Administration. This research made use of APLpy, an open-source plotting package for Python \citep{robitaille12,aplpy2019}.

\section*{Data Availability}

The HST data used for this study are part of the Cycle 25 GO program 15069 and are publicly available on the Barbara A. Mikulski Archive for Space Telescopes (\url{https://archive.stsci.edu/hst/search.php}).



\bibliographystyle{mnras}
\bibliography{dynamo} 




\appendix
\section{Clump Properties} \label{app:measurements}
Table \ref{tab:measurements} lists the: (1) galaxy ID from Table 3 in \citet{green14}, (2) clump ID, (3) and (4) right ascension and declination of the center of each clump, (5) clump 2\,$\times\sigma$ size from the 1D Gaussian fits of their light profiles, as described in \S\ref{subsec: clumpd_ID}, (6), (7), and (8) the AB magnitude of each clump in the F225W, F336W, and F467M filters, as measured in \S\ref{subsec: photometry}, along with the photometric uncertainties.
\input{measurements}

\section{\texttt{Starburst99} Model Comparison} \label{app:model-comp}
\begin{figure*}
    \centering
    \includegraphics[width=\textwidth]{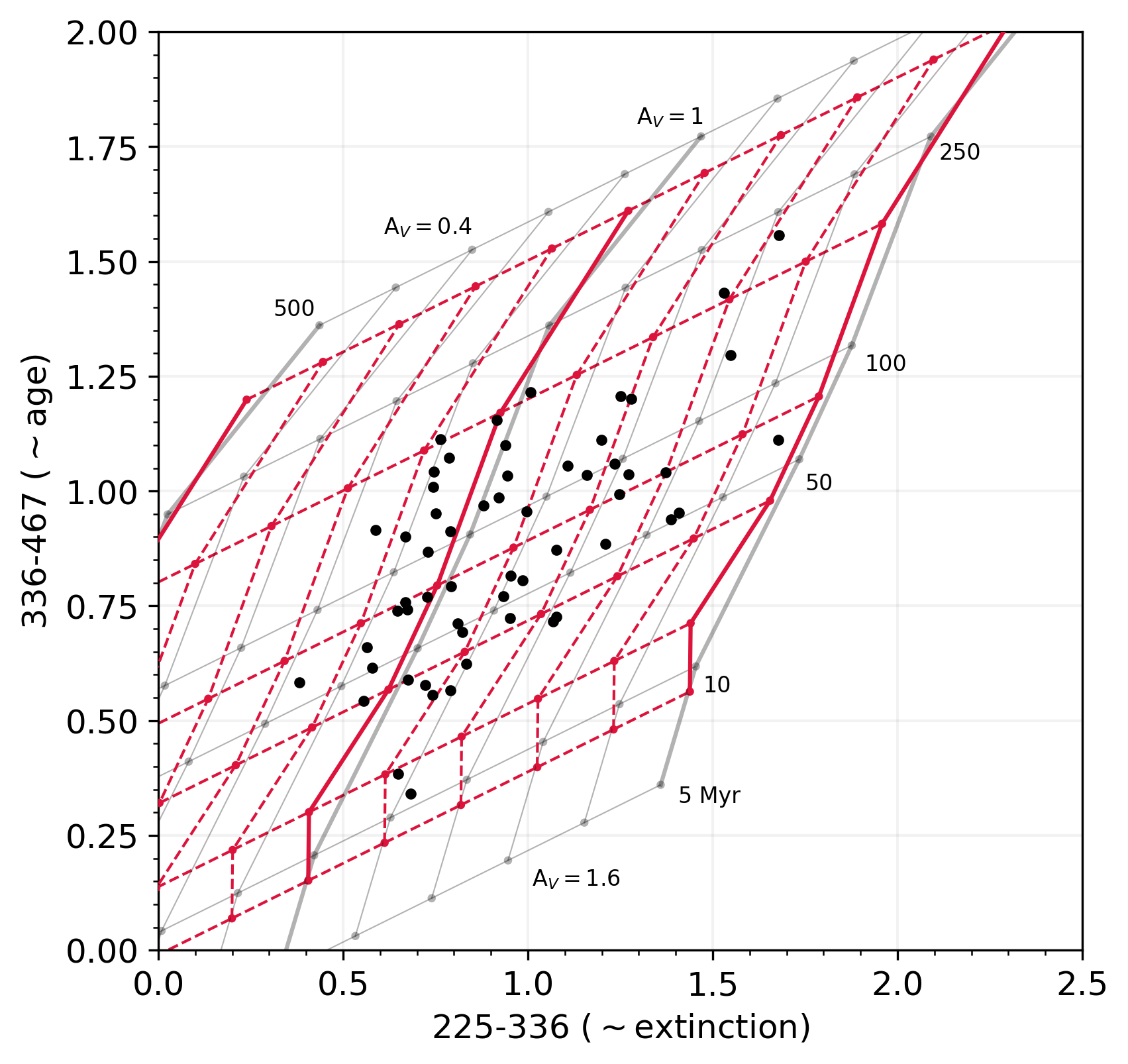}
    \caption{A comparison of \texttt{Starburst99} models: gray lines represent a model with star formation history corresponding to a single burst with solar metallicity, and the red lines represent a model with the same star formation history, but with 40\% solar metallicity. We overlay the clump colours we measure as black circles. The effect that assuming a lower metallicity has on the inferred ages and extinction of the DYNAMO clumps is minimal. The lower metallicity ``compresses'' the range of ages spanned by the $(225 - 336)$ and $(336 - 467 colours)$, such that the youngest two clumps in our sample appear younger in the lower metallicity model. Aside, from these two exceptions, the DYNAMO clumps appear older by a factor of $\sim$\,1.4 in the lower metallicity models.}
    \label{fig:model-comp}
\end{figure*}


\bsp	
\label{lastpage}
\end{document}

%% file: properties.tex
\begin{table*}
	\centering
	\caption{Properties of HST-DYNAMO Galaxies}
	\label{tab:properties}
	\begin{tabular}{lcccccccc} 
		\hline
		Galaxy & $z$ & $M_{*}$ & $M_{gas}$ & $f_{gas}$ & SFR$_{\mathrm{H}_{\alpha}}$ & $<\mathrm{A_{H_{\alpha}}}>$ & $<r_{clump}>$ & 12 + log(O/H) \\
		 &  & (M$_{\odot}$) & (10$^{9}$ M$_{\odot}$) &  & (10$^{9}$ M$_{\odot}$) & (mag) & (pc) & \\
		\hline
		D13-5 & 0.07535 & 53.84 & 20.17 & 0.36 $\pm$ 0.02 & 21.20 & 1.80 $\pm$ 0.52 & 400 & 9.10{\raisebox{0.5ex}{\tiny$^{+9.11}_{-9.09}$}} \\
        D15-3 & 0.06712 & 54.15 & 14.74 & 0.17 $\pm$ 0.02 & 13.70 & $\cdots$ & 394 & 9.18{\raisebox{0.5ex}{\tiny$^{+9.19}_{-9.17}$}} \\
        G04-1 & 0.12981 & 64.64 & 43.68 & 0.33 $\pm$ 0.04 & 41.61 & 1.52 $\pm$ 0.26 & 451 & 9.15{\raisebox{0.5ex}{\tiny$^{+9.16}_{-9.12}$}} \\
        G08-5 & 0.13964 & 9.65 & 16.29 & 0.30 $\pm$ 0.05 & 11.05 & $\cdots$ & 521 & 9.00{\raisebox{0.5ex}{\tiny$^{+9.02}_{-8.98}$}} \\
        G14-1 & 0.13233 & 8.30 & 9.72 & 0.77 $\pm$ 0.08 & 22.33 & $\cdots$ & 353 & 9.10{\raisebox{0.5ex}{\tiny$^{+9.11}_{-9.06}$}} \\
        G20-2 & 0.14113 & 17.27 & 19.12 & 0.21 $\pm$ 0.05 & 21.56 & 0.89 $\pm$ 0.14 & 449 & 8.98{\raisebox{0.5ex}{\tiny$^{+8.99}_{-8.96}$}} \\
		\hline
	\end{tabular}
\end{table*}

%% file: measurements.tex
\begin{table*}
	\centering
	\caption{Properties of HST-DYNAMO Galaxy Clumps}
	\label{tab:measurements}
	\begin{tabular}{llcccccc} 
		\hline
		\hline
		Galaxy & ID & RA & Dec & r$_{clump}$ & m$_{225}$ & m$_{336}$ & m$_{467}$ \\
		 &  & (J2000) & (J2000) & (pc) & (abmag) & (abmag)) & (abmag) \\
		\hline
		
		D13-5 & 1 & 13:30:06.993 & +0:31:54.299 & 389 $\pm$ 14 & 23.78 $\pm$ 0.04 & 22.55 $\pm$ 0.01 & 21.49 $\pm$ 0.03 \\
        D13-5 & 2 & 13:30:06.976 & +0:31:54.115 & 406 $\pm$ 9 & 23.58 $\pm$ 0.03 & 22.48 $\pm$ 0.01 & 21.42 $\pm$ 0.03 \\
        D13-5 & 3 & 13:30:06.961 & +0:31:53.949 & 336 $\pm$ 17 & 24.38 $\pm$ 0.05 & 23.11 $\pm$ 0.02 & 22.07 $\pm$ 0.05 \\
        D13-5 & 4 & 13:30:07.018 & +0:31:54.274 & 256 $\pm$ 7 & 23.85 $\pm$ 0.03 & 22.91 $\pm$ 0.02 & 22.14 $\pm$ 0.04 \\
        D13-5 & 5 & 13:30:06.910 & +0:31:53.361 & 471 $\pm$ 29 & 24.37 $\pm$ 0.10 & 22.82 $\pm$ 0.02 & 21.53 $\pm$ 0.04 \\
        D13-5 & 6 & 13:30:06.930 & +0:31:52.802 & 285 $\pm$ 28 & 25.24 $\pm$ 0.14 & 23.96 $\pm$ 0.06 & 22.76 $\pm$ 0.13 \\
        D13-5 & 7 & 13:30:07.077 & +0:31:53.891 & 410 $\pm$ 22 & 24.04 $\pm$ 0.06 & 22.85 $\pm$ 0.02 & 21.73 $\pm$ 0.06 \\
        D13-5 & 8 & 13:30:07.125 & +0:31:54.052 & 328 $\pm$ 10 &  22.29 $\pm$ 0.01 & 21.61 $\pm$ 0.01 & 21.27 $\pm$ 0.02 \\
        D13-5 & 9 & 13:30:06.985 & +0:31:52.257 & 397 $\pm$ 13 & 23.41 $\pm$ 0.03 & 22.46 $\pm$ 0.02 & 21.64 $\pm$ 0.05 \\
        D13-5 & 10 & 13:30:07.030 & +0:31:52.612 & 402 $\pm$ 37 & 23.79 $\pm$ 0.04 & 22.54 $\pm$ 0.02 & 21.55 $\pm$ 0.04 \\
        D13-5 & 11 & 13:30:07.056 & +0:31:52.641 & 522 $\pm$ 93 & 22.87 $\pm$ 0.02 & 21.71 $\pm$ 0.01 & 20.67 $\pm$ 0.02 \\
        D13-5 & 12 & 13:30:06.999 & +0:31:51.873 & 362 $\pm$ 14 & 23.83 $\pm$ 0.04 & 22.76 $\pm$ 0.02 & 22.03 $\pm$ 0.09 \\
        D13-5 & 13 & 13:30:07.029 & +0:31:52.116 & 422 $\pm$ 11 & 22.42 $\pm$ 0.01 & 21.60 $\pm$ 0.01 & 20.91 $\pm$ 0.02 \\
        D13-5 & 14 & 13:30:07.064 & +0:31:52.308 & 320 $\pm$ 11 & 22.54 $\pm$ 0.01 & 21.75 $\pm$ 0.01 & 21.18 $\pm$ 0.02 \\
        D13-5 & 15 & 13:30:07.052 & +0:31:52.161 & 622 $\pm$ 36 & 21.20 $\pm$ 0.01 & 20.46 $\pm$ 0.004 & 19.90 $\pm$ 0.01 \\
        D13-5 & 16 & 13:30:07.085 & +0:31:52.334 & 459 $\pm$ 19 & 22.69 $\pm$ 0.02 & 21.74 $\pm$ 0.01 & 21.02 $\pm$ 0.02 \\
        D13-5 & 17 & 13:30:07.066 & +0:31:52.096 & 461 $\pm$ 33 & 21.82 $\pm$ 0.01 & 21.17 $\pm$ 0.01 & 20.78 $\pm$ 0.02 \\
        D13-5 & 18 & 13:30:07.088 & +0:31:52.140 & 371 $\pm$ 19 & 23.24 $\pm$ 0.02 & 22.40 $\pm$ 0.01 & 21.78 $\pm$ 0.03 \\
        \hline
        D15-3 & 1 & 15:34:35.452 & -0:28:45.745 & 243 $\pm$ 6 & $<$28.45 & 24.62 $\pm$ 0.04 & 21.83 $\pm$ 0.03 \\
        D15-3 & 2 & 15:34:35.405 & -0:28:46.004 & 373 $\pm$ 16 & 24.75 $\pm$ 0.07 & 23.50 $\pm$ 0.02 & 22.29 $\pm$ 0.05 \\
        D15-3 & 3 & 15:34:35.420 & -0:28:45.217 & 572 $\pm$ 25 & 24.31 $\pm$ 0.12 & 22.78 $\pm$ 0.02 & 21.35 $\pm$ 0.04 \\
        D15-3 & 4 & 15:34:35.327 & -0:28:45.734 & 277 $\pm$ 9 & 25.63 $\pm$ 0.11 & 24.56 $\pm$ 0.06 & 23.69 $\pm$ 0.59 \\
        D15-3 & 5 & 15:34:35.449 & -0:28:43.476 & 398 $\pm$ 22 & 24.76 $\pm$ 0.08 & 23.37 $\pm$ 0.02 & 22.43 $\pm$ 0.08 \\
        D15-3 & 6 & 15:34:35.436 & -0:28:43.562 & 435 $\pm$ 44 & 24.38 $\pm$ 0.07 & 23.01 $\pm$ 0.02 & 21.97 $\pm$ 0.06 \\
        D15-3 & 7 & 15:34:35.308 & -0:28:45.177 & 359 $\pm$ 13 & 24.11 $\pm$ 0.04 & 22.90 $\pm$ 0.02 & 22.01 $\pm$ 0.05 \\
        D15-3 & 8 & 15:34:35.293 & -0:28:45.235 & 389 $\pm$ 32 & 23.67 $\pm$ 0.03 & 22.60 $\pm$ 0.01 & 21.88 $\pm$ 0.04 \\
        D15-3 & 9 & 15:34:35.382 & -0:28:43.822 & 920 $\pm$ 138 & 23.98 $\pm$ 0.20 & 22.30 $\pm$ 0.04 & 20.74 $\pm$ 0.09 \\
        D15-3 & 10 & 15:34:35.374 & -0:28:42.702 & 496 $\pm$ 31 & 23.94 $\pm$ 0.05 & 22.53 $\pm$ 0.02 & 21.58 $\pm$ 0.05 \\
        \hline
        G04-1 & 1 & 4:12:19.650 & -5:54:48.543 & 451 $\pm$ 14 & 23.64 $\pm$ 0.03 & 22.73 $\pm$ 0.01 & 21.57 $\pm$ 0.03 \\
        G04-1 & 2 & 4:12:19.719 & -5:54:47.291 & 321 $\pm$ 11 & 23.73 $\pm$ 0.03 & 22.92 $\pm$ 0.02 & 22.20 $\pm$ 0.05 \\
        G04-1 & 3 & 4:12:19.752 & -5:54:47.164 & 308 $\pm$ 9 & 23.66 $\pm$ 0.03 & 22.94 $\pm$ 0.02 & 22.36 $\pm$ 0.05 \\
        G04-1 & 4 & 4:12:19.703 & -5:54:48.345 & 466 $\pm$ 10 & 22.63 $\pm$ 0.02 & 21.62 $\pm$ 0.01 & 20.41 $\pm$ 0.01 \\
        G04-1 & 5 & 4:12:19.707 & -5:54:48.789 & 487 $\pm$ 11 & 22.49 $\pm$ 0.01 & 21.56 $\pm$ 0.01 & 20.46 $\pm$ 0.01 \\
        G04-1 & 6 & 4:12:19.766 & -5:54:47.759 & 572 $\pm$ 39 & 22.95 $\pm$ 0.02 & 22.01 $\pm$ 0.01 & 20.98 $\pm$ 0.02 \\
        G04-1 & 7 & 4:12:19.745 & -5:54:48.567 & 491 $\pm$ 15 & 22.47 $\pm$ 0.01 & 21.47 $\pm$ 0.01 & 20.52 $\pm$ 0.02 \\
        G04-1 & 8 & 4:12:19.774 & -5:54:48.052 & 366 $\pm$ 23 & 23.80 $\pm$ 0.03 & 22.88 $\pm$ 0.02 & 21.90 $\pm$ 0.04 \\
        G04-1 & 9 & 4:12:19.783 & -5:54:49.985 & 402 $\pm$ 11 & 23.77 $\pm$ 0.03 & 23.21 $\pm$ 0.02 & 22.55 $\pm$ 0.06 \\
        \hline
        G08-5 & 1 & 8:54:18.797 & +6:46:21.327 & 535 $\pm$ 25 & 22.39 $\pm$ 0.01 & 21.81 $\pm$ 0.01 & 21.20 $\pm$ 0.02 \\
        G08-5 & 2 & 8:54:18.798 & +6:46:20.935 & 583 $\pm$ 19 & 22.68 $\pm$ 0.02 & 22.00 $\pm$ 0.01 & 21.26 $\pm$ 0.03 \\
        G08-5 & 3 & 8:54:18.717 & +6:46:19.255 & 335 $\pm$ 9 & 24.27 $\pm$ 0.04 & 23.54 $\pm$ 0.02 & 22.77 $\pm$ 0.06 \\
        G08-5 & 4 & 8:54:18.792 & +6:46:19.894 & 506 $\pm$ 35 & 24.49 $\pm$ 0.08 & 22.81 $\pm$ 0.02 & 21.70 $\pm$ 0.03 \\
        \hline
        G14-1 & 1 & 14:54:28.329 & +0:44:34.584 & 353 $\pm$ 12 & 21.08 $\pm$ 0.01 & 20.41 $\pm$ 0.004 & 19.82 $\pm$ 0.01 \\
        G14-1 & 2 & 14:54:28.314 & +0:44:33.917 & 455 $\pm$ 10 & 22.75 $\pm$ 0.01 & 22.10 $\pm$ 0.01 & 21.36 $\pm$ 0.02 \\
        G14-1 & 3 & 14:54:28.393 & +0:44:34.258 & 262 $\pm$ 5 & 23.37 $\pm$ 0.02 & 22.39 $\pm$ 0.01 & 21.58 $\pm$ 0.02 \\
        \hline
        G20-2 & 1 & 20:44:02.885 & -6:46:57.221 & 452 $\pm$ 20 & 22.72 $\pm$ 0.01 & 22.13 $\pm$ 0.01 & 21.22 $\pm$ 0.02 \\
        G20-2 & 2 & 20:44:02.886 & -6:46:57.506 & 537 $\pm$ 29 & 21.82 $\pm$ 0.01 & 21.15 $\pm$ 0.01 & 20.25 $\pm$ 0.01 \\
        G20-2 & 3 & 20:44:02.900 & -6:46:57.535 & 870 $\pm$ 173 & 21.30 $\pm$ 0.01 & 20.55 $\pm$ 0.004 & 19.51 $\pm$ 0.01 \\
        G20-2 & 4 & 20:44:02.901 & -6:46:57.766 & 612 $\pm$ 53 & 21.37 $\pm$ 0.01 & 20.59 $\pm$ 0.004 & 19.51 $\pm$ 0.01 \\
        G20-2 & 5 & 20:44:02.934 & -6:46:57.367 & 363 $\pm$ 48 & 22.71 $\pm$ 0.02 & 21.95 $\pm$ 0.01 & 20.84 $\pm$ 0.02 \\
        G20-2 & 6 & 20:44:02.911 & -6:46:58.218 & 454 $\pm$ 41 & 22.46 $\pm$ 0.01 & 21.72 $\pm$ 0.01 & 20.71 $\pm$ 0.02 \\
        G20-2 & 7 & 20:44:02.955 & -6:46:57.455 & 491 $\pm$ 23 & 21.87 $\pm$ 0.01 & 21.12 $\pm$ 0.01 & 20.17 $\pm$ 0.01 \\
        G20-2 & 8 & 20:44:02.969 & -6:46:57.253 & 425 $\pm$ 20 & 22.43 $\pm$ 0.01 & 21.76 $\pm$ 0.01 & 21.01 $\pm$ 0.02 \\
        G20-2 & 9 & 20:44:02.928 & -6:46:58.409 & 446 $\pm$ 25 & 22.69 $\pm$ 0.02 & 21.90 $\pm$ 0.01 & 20.99 $\pm$ 0.02 \\
        G20-2 & 10 & 20:44:02.949 & -6:46:58.266 & 352 $\pm$ 26 & 22.99 $\pm$ 0.02 & 22.27 $\pm$ 0.01 & 21.40 $\pm$ 0.02 \\
        G20-2 & 11 & 20:44:02.980 & -6:46:57.528 & 669 $\pm$ 47 & 21.62 $\pm$ 0.01 & 20.82 $\pm$ 0.01 & 20.03 $\pm$ 0.01 \\
        G20-2 & 12 & 20:44:02.964 & -6:46:58.029 & 354 $\pm$ 42 & 23.23 $\pm$ 0.02 & 22.35 $\pm$ 0.01 & 21.38 $\pm$ 0.03 \\
        G20-2 & 13 & 20:44:03.016 & -6:46:57.556 & 331 $\pm$ 16 & 23.87 $\pm$ 0.03 & 23.32 $\pm$ 0.02 & 22.77 $\pm$ 0.09 \\
        G20-2 & 14 & 20:44:02.990 & -6:46:58.564 & 401 $\pm$ 18 & 23.02 $\pm$ 0.02 & 22.64 $\pm$ 0.01 & 22.06 $\pm$ 0.04 \\
        \hline
	\end{tabular}
\end{table*}